\documentclass[aps, prb, twocolumn]{revtex4-1}

\usepackage{enumerate}
\usepackage{comment}
\usepackage{amsmath}
\usepackage{amssymb}
\usepackage{color}
\usepackage[dvipsnames]{xcolor}
\usepackage{tikz}
\usepackage{amsthm}
\usepackage{graphicx}
\usepackage{pict2e}
\usepackage{bbold}
\usepackage{empheq}
\usepackage{subfigure}
\usepackage{xfrac}
\newcommand{\be}{\begin{equation}}
\newcommand{\ee}{\end{equation}}
\newcommand{\bea}{\begin{eqnarray}}
\newcommand{\eea}{\end{eqnarray}}
\newcommand{\beb}{\begin{eqnarray*}}
\newcommand{\eeb}{\end{eqnarray*}}

\newcommand{\LD}{\langle}
\newcommand{\RD}{\rangle}

\newcommand{\dotc}{\text{d}^{\dagger}}
\newcommand{\dota}{\text{d}}
\newcommand{\leadc}{\text{c}^{\dagger}}
\newcommand{\leada}{\text{c}}

\newcommand{\app}{Appendix }
\newcommand{\sect}{Sec.~}
\newcommand{\eqn}{Eq.~}
\newcommand{\eqns}{Eqs.~}
\newcommand{\fig}{Fig.~}
\newcommand{\figs}{Figs.~}
\newcommand{\reff}{Ref.~}

\newcommand{\SPState}{\Psi }
\newcommand{\SPstate}{\psi }

\newcommand{\TP}{two-particle}
\newcommand{\SP}{single-particle}
\newcommand{\SO}{spin-orbit}

\newcommand{\HTun}{\text{H}_{T}}
\newcommand{\HLeads}{\text{H}_{leads}}
\newcommand{\Ecap}{E_{c}}
\newcommand{\CapC}{C_{G}}
\newcommand{\Vgate}{V_{G}}
\newcommand{\Vbias}{V_{B}}

\newcommand{\TPEn}{\mathfrak{E}}

\newcommand{\BLG}{bilayer graphene}

\newcommand{\GS}{ground state}

\newcommand{\RN}[1]{%
  \textup{\uppercase\expandafter{\romannumeral#1}}%
}

\newcommand{\QD}{quantum dot}
\newcommand{\QPC}{quantum point contact}

\newcommand{\SPspin}{\mathfrak{s}}
\newcommand{\SPvalley}{\mathfrak{t}}
\newcommand{\SPorbital}{n}
\newcommand{\SPorbitalm}{m}

\newcommand{\TPsymm}{s}
\newcommand{\TPasymm}{a}

\newcommand{\gVTPs}{2g_{v}^{\SPorbital}}
\newcommand{\gVTPa}{(g_{v}^{\SPorbital}+g_{v}^{\SPorbitalm})}
\newcommand{\gVSP}{g_{v}^{\SPorbital}}

\newcommand{\downplus}{\downarrow +}
\newcommand{\downminus}{\downarrow -}
\newcommand{\upplus}{\uparrow +}
\newcommand{\upminus}{\uparrow -}

\newcommand{\partnumb}{\mathcal{N}}

\newcommand{\partnumbcol}{\partnumb\!:}
\newcommand{\dotstate}{\mathcal{N}\!:\chi}
\newcommand{\dotstateP}{\mathcal{N}^{\prime}\!:\chi^{\prime}}
\newcommand{\dotprob}{\text{P}}

\newcommand{\broad}{\Gamma}
\newcommand{\Chempot}{\mu}
\newcommand{\intW}{\mathfrak{J}}

\newcommand{\LnnSzeroTplusX}{\SPorbital\SPorbital, {\sigma^{-x},\tau^{+x}}}
\newcommand{\LnnSzeroTminusZ}{\SPorbital\SPorbital, { \sigma^{-x},\tau^{-z}}}
\newcommand{\LnnSzeroTplusZ}{ \SPorbital\SPorbital,{ \sigma^{-x},\tau^{+z}}}
\newcommand{\LnnmminusOneTminusX}{\SPorbital\SPorbital, {\sigma^{-z},\tau^{-x}}}
\newcommand{\LnnmplusOneTminusX}{\SPorbital\SPorbital, {\sigma^{+z},\tau^{-x}}}
\newcommand{\LnnmZeroTminusX}{ \SPorbital\SPorbital,{\sigma^{+x},\tau^{-x}}}

\newcommand{\StatenSzeroTplusX}{|\SPorbital\SPorbital,\sigma^{-x},\tau^{+x}\rangle}
\newcommand{\StatenSzeroTminusZ}{|\SPorbital\SPorbital,\sigma^{-x},\tau^{-z}\rangle}
\newcommand{\StatenSzeroTplusZ}{|\SPorbital\SPorbital,\sigma^{-x},\tau^{+z}\rangle}
\newcommand{\StatenmminusOneTminusX}{|\SPorbital\SPorbital,\sigma^{-z},\tau^{-x}\rangle}
\newcommand{\StatenmplusOneTminusX}{|\SPorbital\SPorbital,\sigma^{+z},\tau^{-x}\rangle}
\newcommand{\StatenmZeroTminusX}{|\SPorbital\SPorbital,\sigma^{+x},\tau^{-x}\rangle}

\newcommand{\StatenmmminusOneTminusZ}{|\SPorbital\SPorbitalm,\;\sigma^{-z},\;\tau^{-z}\rangle}
\newcommand{\StatenmmZeroTminusZ}{|\SPorbital\SPorbitalm,\;\sigma^{+x},\;\tau^{-z}\rangle}
\newcommand{\StatenmmplusOneTminusZ}{|\SPorbital\SPorbitalm,\;\sigma^{+z},\;\tau^{-z}\rangle}
\newcommand{\StatenmmminusOneTplusX}{|\SPorbital\SPorbitalm,\;\sigma^{-z},\;\tau^{+x}\rangle}
\newcommand{\StatenmmZeroTplusX}{|\SPorbital\SPorbitalm,\;\sigma^{+x},\;\tau^{+x}\rangle}
\newcommand{\StatenmmplusOneTplusX}{|\SPorbital\SPorbitalm,\;\sigma^{+z},\;\tau^{+x}\rangle}
\newcommand{\StatenmmminusOneTplusZ}{|\SPorbital\SPorbitalm,\;\sigma^{-z},\;\tau^{+z}\rangle}
\newcommand{\StatenmmZeroTplusZ}{|\SPorbital\SPorbitalm,\;\sigma^{+x},\;\tau^{+z}\rangle}
\newcommand{\StatenmmplusOneTplusZ}{|\SPorbital\SPorbitalm,\;\sigma^{+z},\;\tau^{+z}\rangle}
\newcommand{\StatenmSzeroTminusX}{|\SPorbital\SPorbitalm,\;\sigma^{-x},\;\tau^{-x}\rangle}

\newcommand{\LnmmminusOneTminusZ}{\SPorbital\SPorbitalm, {\sigma^{-z}, \tau^{-z}}}
\newcommand{\LnmmZeroTminusZ}{ \SPorbital\SPorbitalm, {\sigma^{+x}, \tau^{-z}}}
\newcommand{\LnmmplusOneTminusZ}{ {\SPorbital\SPorbitalm, \sigma^{+z}, \tau^{-z}}}
\newcommand{\LnmmminusOneTplusX}{ {\SPorbital\SPorbitalm, \sigma^{-z}, \tau^{+x}}}
\newcommand{\LnmmZeroTplusX}{\SPorbital\SPorbitalm, {\sigma^{+x}, \tau^{+x}}}
\newcommand{\LnmmplusOneTplusX}{\SPorbital\SPorbitalm, {\sigma^{+z}, \tau^{+x}}}
\newcommand{\LnmmminusOneTplusZ}{\SPorbital\SPorbitalm, {\sigma^{-z}, \tau^{+z}}}
\newcommand{\LnmmZeroTplusZ}{\SPorbital\SPorbitalm, {\sigma^{+x}, \tau^{+z}}}
\newcommand{\LnmmplusOneTplusZ}{\SPorbital\SPorbitalm, {\sigma^{+z}, \tau^{+z}}}
\newcommand{\LnmSzeroTminusX}{\SPorbital\SPorbitalm, {\sigma^{-x}, \tau^{-x}}}

\begin{document}

\title{Theory of tunneling spectra for a few-electron bilayer graphene quantum dot}

\author{Angelika Knothe$^{1}$}
\author{Leonid I.~Glazman$^{2}$}
\author{Vladimir I.~Fal'ko$^{1,3,4}$}
\affiliation{$^1$National Graphene Institute, University of Manchester, Manchester M13 9PL, United Kingdom}
\affiliation{$^2$Departments of Physics and Applied Physics, Yale University, New Haven, CT 06520, USA}
\affiliation{$^3$Department of Physics and Astronomy, University of Manchester, Oxford Road, Manchester, M13 9PL, United Kingdom}
\affiliation{$^4$Henry Royce Institute for Advanced Materials, University of Manchester, Manchester, M13 9PL, United Kingdom}
\date{\today}

\begin{abstract}
{The tuneability and control of quantum nanostructures in two-dimensional materials offer promising perspectives for their use in future electronics. It is hence necessary to analyze quantum transport in such nanostructures. Material properties such as a complex dispersion, topology, and charge carriers with multiple degrees of freedom, are appealing for novel device functionalities but complicate their theoretical description. Here, we study quantum tunnelling transport across a few-electron bilayer graphene quantum dot. We demonstrate how to uniquely identify single- and two-electron dot states' orbital, spin, and valley composition from differential conductance in a finite magnetic field. Furthermore, we show that the transport features manifest splittings in the dot's spin and valley multiplets induced by interactions and magnetic field (the latter splittings being a consequence of bilayer graphene's Berry curvature). Our results elucidate spin- and valley-dependent tunnelling mechanisms and will help to utilize bilayer graphene quantum dots, e.g., as spin and valley qubits.}
\end{abstract}
\maketitle

\section{Introduction}

{Carbon-based materials are considered promising candidates for spin-based quantum computation devices due to their low spin-orbit and hyperfine coupling entailing long spin coherence life times \cite{drogelerSpinLifetimesExceeding2016, ingla-aynesEightyEightPercentDirectional2016, avsarColloquiumSpintronicsGraphene2020}. Any spin-qubit operation using a \QD{} will necessarily include the steps of controlled  {loading} (transferring a charge carrier onto the dot) and  {storage} (keeping the charge carrier on the dot). 
Such an operation hence requires understanding and control of the dot's few-electron states and tunnel transport processes.} 

\begin{figure}[htb]
 \centering
\includegraphics[width=1\linewidth]{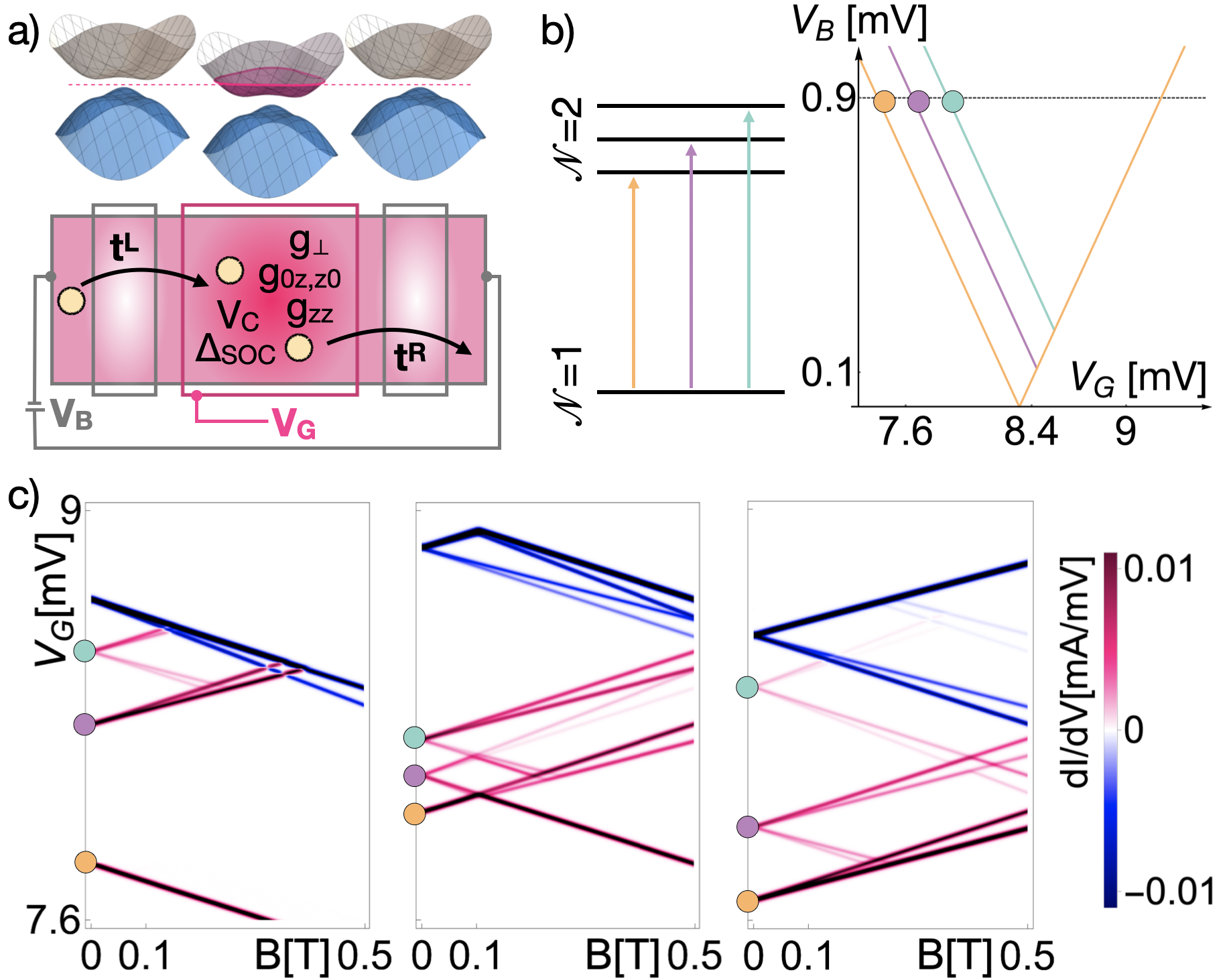}
\caption{ {a) Bilayer graphene lead-dot-lead setup. b) Single ($\mathcal{N}=1$) to \TP{} ($\mathcal{N}=2$) tunnelling transitions allow characterising the dot's  orbital, spin and valley states. The \GS{} transition defines the Coulomb diamond in the bias ($V_{B}$) and gate voltage ($V_{G}$) plane. Transitions contribute to transport depending on spin and valley  selection rules. c) The slope of differential conductance lines with the magnetic field (dominated by the difference of the \TP{} and \SP{} valley g-factors) teaches about the orbital and valley composition of the \TP{} dot state. The splittings between transitions at $B=0$ manifest the interaction-induced \TP{} state gaps.}}
\label{fig:Teaser}
\end{figure}

 In \BLG{},  recent experiments achieve confinement of charge carriers in one- and zero-dimensional structures by electrostatic gating \cite{droscherElectronFlowSplitgated2012, overwegElectrostaticallyInducedQuantum2018, overwegTopologicallyNontrivialValley2018, kraftValleySubbandSplitting2018, banszerusGateDefinedElectronHole2018, geVisualizationManipulationBilayer2020, banszerusObservationSpinOrbitGap2020}. To electrostatically define a nanostructure in \BLG{} multiple gates locally modulate the \BLG{} band gap and charge carrier density, cf.~\fig\ref{fig:Teaser}a). Split gates can define a channel (pink strip in  \fig\ref{fig:Teaser}a)), while  finger gates on top create a dot-like region within this channel (dark pink region), bounded by gapped regions acting as barriers  (white regions). This confinement method offers immense gate-control of the nanostructure, e.g., the confinement width, depth, barriers, and \BLG{} gap. It is now possible to operate such an electrostatically confined \BLG{} dot controllably in the single and few-electron regime \cite{eichSpinValleyStates2018, eichCoupledQuantumDots2018, banszerusSingleElectronDoubleQuantum2020, banszerusElectronHoleCrossover2020, banszerusDispersiveSensingCharge2020, banszerusPulsedgateSpectroscopySingleelectron2020, banszerusSpinvalleyCouplingSingleelectron2021, kurzmannKondoEffectSpinorbit2021, garreisShellFillingTrigonal2020}. 
  The rapid experimental progress in device design, quality, and control, calls for a theoretical investigation of single and few-electron tunnelling processes in such structures. 
\begin{figure}[t!]
\centering
\begin{picture}(260,230)
\put(0,0){\includegraphics[width=1\linewidth]{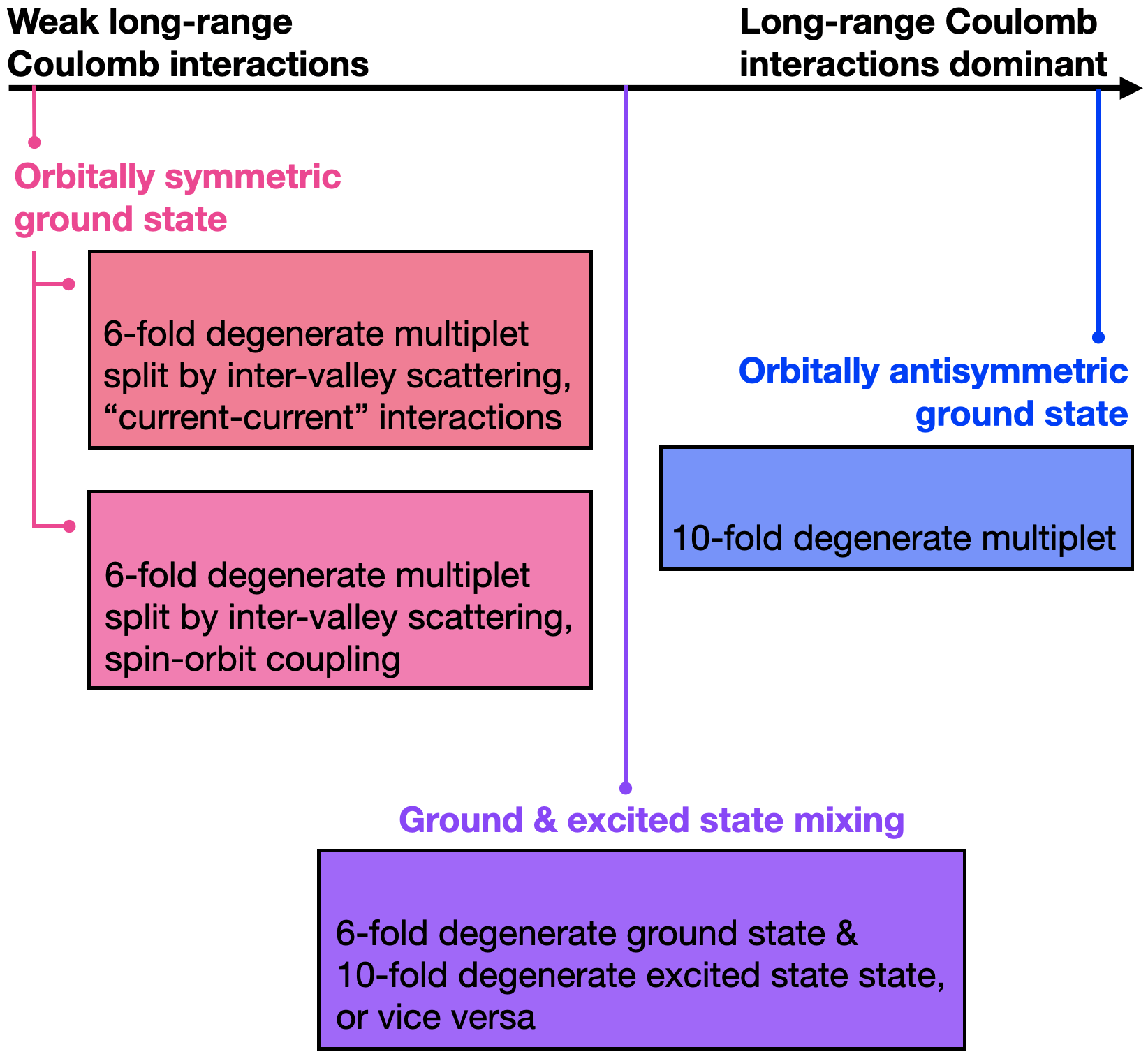}}
\put(21,111){{\fontfamily{phv}\selectfont (\ref{sec:S-TRS})}}
\put(21,162){{\fontfamily{phv}\selectfont (\ref{sec:S-TRSB})} }
\put(218,120){{\fontfamily{phv}\selectfont(\ref{sec:AS})}}
\put(123,34){{\fontfamily{phv}\selectfont(\ref{sec:Cross})}}
\end{picture}
	\caption{We consider transport through \BLG{} \QD{}s in  regimes dominated by different types of interactions. Weak or strong long-range Coulomb interactions favour  \TP{} states with symmetric or antisymmetric orbital states and distinct degeneracies of the spin and valley multiplets. Possible short-range interaction mechanisms include couplings generated by inter-valley scattering or ''current-current'' interactions, and \SO{} coupling. In brackets, we indicate the corresponding section number.}
	\label{fig:Structure}
\end{figure}
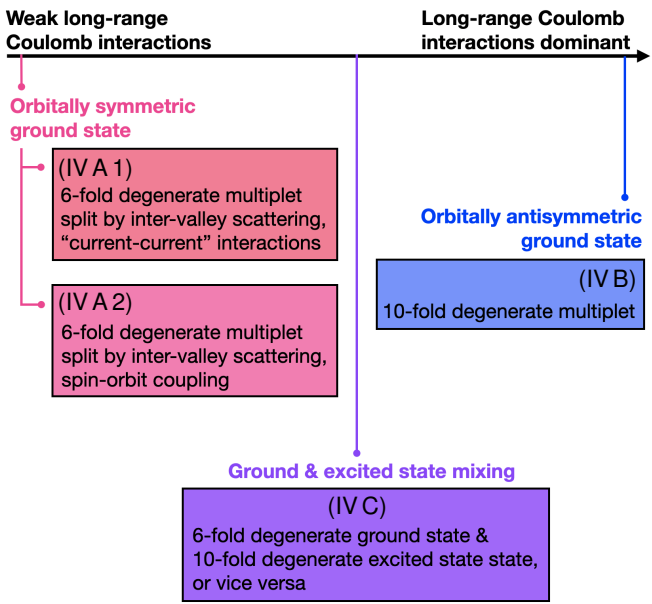

 The two internal degrees of freedom, valley and spin, enrich the spectra of \BLG{}-based devices. The result is highly degenerate multiplets split in various ways by a magnetic field and weak perturbations.
{In this work, we investigate tunnelling transport through a \BLG{} \QD{} in the single and few-electron regime as a tool to unravel some of the dot's \TP{} states' unusual characteristics. We demonstrate how the specifics of the dot's multiplets manifest in tunnelling current, cf.~\fig\ref{fig:Teaser}c), and how to link such experimentally observable transport features with interaction and field-induced gaps between different spin and valley configurations. We determine the particular tunnelling sequences for spin and valley states of differently ordered multiplets and relate them to microscopic parameters, such as short-range interaction coupling constants, $g_{\perp}, g_{zz}, g_{0z}, g_{z0}$\cite{lemonikSpontaneousSymmetryBreaking2010, lemonikCompetingNematicAntiferromagnetic2012, kharitonovPhaseDiagramNu2012, knotheQuartetStatesTwoelectron2020}, and topological valley g-factors (the latter induced by Berry-curvature\cite{xiaoBerryPhaseEffects2010, moulsdaleEngineeringTopologicalMagnetic2020, parkValleyFilteringDue2017, Fuchs2010}). Besides spin- and valley selection rules, these tunnelling sequences depend on the dot-lead coupling characteristics, such as asymmetric coupling to the source and drain and cotunnelling corrections. By combining the aspects of state multiplicity, electronic interactions, and dot-lead couplings, our results add to the understanding of tunnelling transport in complex few-electron systems.} 

The paper is structured as follows. In \sect\ref{sec:Model}, we introduce our theoretical model of the \BLG{} \QD{} and the leads, discussing the dot's state structure in the single- and \TP{} sector. Section \ref{sec:Tunnel} describes the rates for tunnelling between these states and the leads, and the calculation of tunnel current using rate equations. Section \ref{sec:TPStates}  presents our calculations of tunnel transport through a \BLG{} \QD{}. We provide maps of the differential conductance, $dI/dV$, in the plane spanned by the gate voltage and the magnetic field for representative cases of interaction parameters. This way, we characterise  regimes in which different electronic interactions dominate, as tabulated  in \fig\ref{fig:Structure}. 
The differential conductance in a proper bias interval reveals the transitions between the one- and \TP{} states in the \QD{}. Levels energies are closely related to the symmetries of the corresponding orbital wave functions. The multiplicity and ordering of the  \TP{}  levels depends on the orbital symmetry, the short-range part of interactions, and the external magnetic field. The latter allows one to affect the level ordering. The interpretation of such tunnelling data may depend on device characteristics, e.g., the lead-dot coupling strength or uniformity of source and drain coupling. Taking these device features into account, we show how to use the differential conductance maps to identify the dot's \TP{} \GS{} and determine the dominant microscopic interaction parameters. 
Section \ref{sec:conc} contains step-by-step instructions how to use our results to interpret differential conductance data for identifying the single-and \TP{} state structure of a \BLG{} \QD{}.



\section{Model}
\label{sec:Model}

We consider a lead-dot-lead setup in which a \BLG{} \QD{} is tunnel-coupled to \BLG{} \QPC{}s as in \fig\ref{fig:Teaser}a).

\textbf{Single-particle states of the \BLG{} \QD{}.} 
We focus on the experimentally accessible regime of small and moderate displacement fields in the dot region. For a small gap, the \BLG{} dispersion in the vicinity of the $K$-points is approximately quadratic, and a \QD{}'s \SP{} level structure resembles that of harmonic confinement, featuring an orbitally singly degenerate \GS{}\cite{knotheQuartetStatesTwoelectron2020}.
These \SP{} dot states are characterized by the orbital quantum number, $\SPorbital$, and the electron's spin ($\SPspin=\uparrow, \downarrow$) and valley ($\SPvalley=+,-$) degree of freedom. We denote a one-electron dot state by $|\SPorbital,\SPspin,\SPvalley\rangle=\dotc_{\SPorbital \SPspin \SPvalley}|0\rangle$, where $\dotc_{\SPorbital \SPspin \SPvalley}$ is the electron creation operator and $|0\rangle$ is the empty dot state. The $\SPorbital$-th spin and valley multiplet at zero magnetic field is characterized by energy, $E_{\SPorbital}$. Zero-point vibrations enhance Kane-Mele spin-orbit coupling\cite{ochoaSpinorbitCouplingAssisted2012}, $\Delta_{SO}$, leading to reversed spin splitting in opposite two valleys.  Each multiplet splits upon the application of a magnetic field, $B$, perpendicular to the \BLG{} plane as,
\begin{equation}
E_{\SPorbital,\SPspin=\uparrow, \downarrow,\SPvalley}=E_{\SPorbital}+ \Ecap(1)\pm\SPvalley\Delta_{SO}\pm \frac{1}{2} g\mu_{B} B +\SPvalley \gVSP\mu_{B} B,
\label{eqn:SPEn}
\end{equation}
according to the free electron spin g-factor, $g=2$, and valley g-factor, $\gVSP$ ($\mu_{B}$ being the Bohr magneton). The latter is a consequence of gapped \BLG{}'s nontrivial Bloch band Berry curvature entailing an topological orbital magnetic moment  with opposite sign in the two different valleys \cite{xiaoBerryPhaseEffects2010, moulsdaleEngineeringTopologicalMagnetic2020, parkValleyFilteringDue2017, Fuchs2010}. 
As the orbital magnetic moment is a function of wave number in each valley, the topological valley g-factor depends on the gap and the states' momentum space distribution (and, consequently, on the orbital quantum number, $\SPorbital$), determining how much orbital magnetic moment is picked up by the dot states\cite{knotheInfluenceMinivalleysBerry2018, knotheQuartetStatesTwoelectron2020, leeTunableValleySplitting2020, tongTunableValleySplitting2020}. The second term in \eqn\eqref{eqn:SPEn} accounts for the presence of a gate with capacitance $\CapC$, which, at gate voltage $\Vgate$, induces an effective charge on the dot, changing the dot's electrostatic potential by 
\begin{equation}
\Ecap(\partnumb)=\frac{(\partnumb e - \CapC\;  \Vgate)^{2}}{2e C}.
\label{eqn:Ecap}
\end{equation}
 Here, $\partnumb$ is the dot occupation number and $C$ is the total capacitance of the dot.

\textbf{Two-particle states of the \BLG{} \QD{}.}
The dot's \TP{} sector is non-trivial, due to the large number of states arising from different combinations of the orbital, spin, and valley degrees of freedom. Moreover, these degrees of freedom are not independent since all three combined must form an overall antisymmetric \TP{} wave function. 
As we showed in \reff\onlinecite{knotheQuartetStatesTwoelectron2020},  Coulomb interaction between the two dot electrons further impacts the correlations between the different degrees of freedom.

 The long-range Coulomb interaction on the scale of the dot state wave functions is given by,
\begin{align}
\nonumber&H_{C}={\frac{1}{2}\!\iint \!d\mathbf{r}d\mathbf{r}^{\prime}[\SPState^{{\dagger}}_{\SPorbital}(\mathbf{r})\SPState_{\SPorbital}(\mathbf{r})] \, V_{C}(\mathbf{r}-\mathbf{r}^{\prime})\,[\SPState^{{\dagger}}_{\SPorbital}(\mathbf{r}^{\prime})\SPState_{\SPorbital}(\mathbf{r}^{\prime})]},\\
\nonumber&\SPState_{\SPorbital}(\mathbf{r})=\\
&(\SPstate_{\SPorbital\upplus}^{A}, \SPstate_{\SPorbital\upplus}^{B^{\prime}}, \SPstate_{\SPorbital\upminus}^{B^{\prime}}, - \SPstate_{\SPorbital\upminus}^{A}, \SPstate_{\SPorbital\downplus}^{A}, \SPstate_{\SPorbital\downplus}^{B^{\prime}}, \SPstate_{\SPorbital\downminus}^{B^{\prime}}, - \SPstate_{\SPorbital\downminus}^{A})^{T},
\label{eqn:HC}
\end{align}
between the  low-energy electronic fields,  $\SPState_{\SPorbital}(\mathbf{r})$, on the non-dimer sites  $A$ and $B^{\prime}$ of the \BLG{} lattice. We employ the  2D screened Coulomb interaction in a weakly gapped \BLG{} \cite{cheianovGappedBilayerGraphene2012, knotheQuartetStatesTwoelectron2020}, with Fourier representation, $V_{C}(\mathbf{q})=\frac{e^2}{4\pi\epsilon_0\epsilon}\frac{2\pi}{q(1+q R_{\star})}$, where  $\epsilon_0$ is the vacuum permittivity, $\epsilon $ is the encapsulating substrate material's dielectric constant, $R_{\star}=  \sqrt{32} \hbar\kappa /\sqrt{\Delta}$, taking into account gapped \BLG{}'s polarisability\cite{cheianovGappedBilayerGraphene2012}, $\kappa^2 =2m e^2/(4\pi\epsilon_0\epsilon \hbar \sqrt{\Delta} )^2  $, with $m$ being the effective mass and $\Delta$ the \BLG{} gap. 
The Coulomb repulsion in \eqn\eqref{eqn:HC} determines the spatial extent of the wave functions and the exchange energy. The competition of \SP{} energies, direct-, and exchange-interaction terms determines the mixing of \SP{} orbitals forming orbitally symmetric or antisymmetric \TP{} states\cite{knotheQuartetStatesTwoelectron2020}. For zero or weak Coulomb interaction (strong screening by the surrounding medium), two electrons on the dot form an orbitally symmetric wave function, both occupying the same \SP{} orbital, $\SPorbital$. If the Coulomb repulsion dominates (weak screening), the gain in exchange energy overcomes the cost of occupying higher \SP{} orbitals, and the \TP{} \GS{} forms an antisymmetric orbital wave function involving different \SP{} orbitals, $\SPorbital$ and $\SPorbitalm$.

In gapped \BLG{}, where the gating needed to form the \QD{} lifts the  layer symmetry, we take into account the lattice-scale symmetry breaking short-range interactions\cite{lemonikSpontaneousSymmetryBreaking2010, lemonikCompetingNematicAntiferromagnetic2012, kharitonovPhaseDiagramNu2012, knotheQuartetStatesTwoelectron2020},
\begin{equation}
H_{SR}=\frac{1}{2}\int d\mathbf{r}\sum_{{(i,j)}} g_{ij}[\SPState^{{\dagger}}_{\SPorbital}(\mathbf{r})  \varsigma^{AB^{\prime}}_{i}\varsigma_{j}^{+-}\SPState_{\SPorbital}(\mathbf{r}) ]^{2},
\label{eqn:HSR}
\end{equation}
with $ \varsigma^{AB^{\prime}}_{i}$ ($\varsigma_{i}^{+-}$) the Pauli matrices in sub-lattice (valley) space and $(i,j)=(xx,xy,yx,yy,zz,z0,0z)$.  The interactions in \eqn\eqref{eqn:HSR} originate from  symmetry breaking fluctuations and the relevant coupling constants 
\begin{equation}
 g_{xx}=g_{xy}=g_{yx}=g_{yy}\equiv g_{\perp},\;\;\;g_{zz}, \;g_{z0}, \;g_{0z},
\label{eqn:gs}
\end{equation}
 favour states with spontaneously broken symmetries\cite{lemonikSpontaneousSymmetryBreaking2010, lemonikCompetingNematicAntiferromagnetic2012}. Inter-valley scattering introduces the coupling  $g_{\perp}$. The couplings $g_{0z,z0}$ correspond to ''current-current'' interactions\cite{aleinerSpontaneousSymmetryBreaking2007}, favouring states with spontaneously broken time-reversal invariance\cite{lemonikCompetingNematicAntiferromagnetic2012}. The case $i=j=0$ is already included in  \eqn\eqref{eqn:HC}. Other possible combination of indices $i,j$ not listed in \eqn\eqref{eqn:gs} do not affect the states in gapped \BLG{} since the corresponding fluctuations are suppressed by the layer polarization.
 
The short-range interactions in \eqn\eqref{eqn:HSR}  introduce anisotropies in the sublattice and valley space  for \TP{} states with  symmetric orbital wave function.  For orbitally antisymmetric \TP{} wave functions,  contact interactions as in  \eqn\eqref{eqn:HSR} are not relevant due to vanishing electronic density at small inter-particle distances. 
Short-range interaction induced splittings hence provide a way to distinguish orbitally symmetric and antisymmetric dot states. 

Any theoretical estimation of the couplings' numerical values comes with inherent uncertainty since they depend on the relevant energy scale. The resulting renormalization and additional phonon-mediated effects can change the couplings $g_{ij}$ in absolute value and sign\cite{lemonikCompetingNematicAntiferromagnetic2012, kharitonovPhaseDiagramNu2012}. By studying tunnelling through \TP{} multiplets for all possible combinations of values in \eqn\eqref{eqn:gs} we demonstrate how to identify different parameters in transport. Our results will be relevant for unfolding experimental measurements using tunnelling spectroscopy of the \BLG{} \QD{}'s \TP{} states as a tool to extract the microscopic short-range interaction parameters in \eqn\eqref{eqn:gs}.

Depending on the symmetry of the \TP{} states' orbital part, any combination of spin and valley states is permissible that combines to an overall antisymmetric  \TP{} wave function.
There are six combinations of spin/valley-singlet ($\sigma^{-x}/\tau^{-x}$) and -triplet ($\sigma^{-z}/\tau^{-z},\;\sigma^{+x}/\tau^{+x},\;\sigma^{+z}/\tau^{+z}$) states and an orbitally symmetric (\TPsymm) \TP{} state of orbital $\SPorbital$:
\begin{align}
&\nonumber\StatenSzeroTplusX=\frac{1}{\sqrt{2}}(\dotc_{\SPorbital\upplus}\dotc_{\SPorbital\downminus}-\dotc_{\SPorbital\downplus}\dotc_{\SPorbital\upminus})|0\rangle,\\
&\nonumber\StatenSzeroTminusZ=\dotc_{\SPorbital\upminus}\dotc_{\SPorbital\downminus}|0\rangle,\\
&\nonumber\StatenSzeroTplusZ=\dotc_{\SPorbital\upplus}\dotc_{\SPorbital\downplus}|0\rangle,\\
&\nonumber\StatenmminusOneTminusX=\dotc_{\SPorbital\downplus}\dotc_{\SPorbital\downminus}|0\rangle,\\
&\nonumber\StatenmZeroTminusX=\frac{1}{\sqrt{2}}(\dotc_{\SPorbital\upplus}\dotc_{\SPorbital\downminus}+\dotc_{\SPorbital\downplus}\dotc_{\SPorbital\upminus})|0\rangle,\\
&\StatenmplusOneTminusX=\dotc_{\SPorbital\upplus}\dotc_{\SPorbital\upminus}|0\rangle.
\label{eqn:sStates}
\end{align}
The energies of this \TP{}  multiplet 
are given by\cite{knotheQuartetStatesTwoelectron2020}, 
\begin{align}
\nonumber &E_{\LnnSzeroTplusX}\\ \nonumber&=\TPEn^{\TPsymm}_{\SPorbital\SPorbital} +(g_{zz}+ 4g_{\perp}-g_{0z}-g_{z0})\intW  + \Ecap(2) ,\\
\nonumber &E_{\LnnSzeroTminusZ}\\ \nonumber&=\TPEn^{\TPsymm}_{ \SPorbital\SPorbital} + (g_{zz} +g_{0z}+g_{z0})\intW + \Ecap(2) -\gVTPs \mu_{B} B,\\
\nonumber &E_{\LnnSzeroTplusZ}\\ \nonumber&=\TPEn^{\TPsymm}_{ \SPorbital\SPorbital} + (g_{zz} +g_{0z}+g_{z0})\intW + \Ecap(2) +\gVTPs\mu_{B} B,\\
\nonumber &E_{\LnnmminusOneTminusX}\\ \nonumber&=\TPEn^{\TPsymm}_{ \SPorbital\SPorbital} + (g_{zz}-4g_{\perp}-g_{0z}-g_{z0})\intW + \Ecap(2) -g\mu_{B} B,\\
\nonumber &E_{\LnnmZeroTminusX}\\ \nonumber&=\TPEn^{\TPsymm}_{ \SPorbital\SPorbital} + (g_{zz}-4g_{\perp}-g_{0z}-g_{z0})\intW + \Ecap(2) ,\\
\nonumber& E_{\LnnmplusOneTminusX}\\ &=\TPEn^{\TPsymm}_{ \SPorbital\SPorbital} + (g_{zz}-4g_{\perp}-g_{0z}-g_{z0})\intW + \Ecap(2) +g \mu_{B}B.
 \label{eqn:sEnergies}
\end{align}
 Here,  $\TPEn^{\TPsymm}_{ \SPorbital\SPorbital}$, comprises the energy of the $\SPorbital$-th \SP{} orbital and the screened electron-electron Coulomb interaction computed from \eqn\eqref{eqn:HC}. 
The factor, $\intW=\int d\mathbf{r}[\psi_{\SPorbital\SPspin_1\SPvalley_1}^{B^{\prime}}(\mathbf{r})]^*[\psi_{\SPorbital\SPspin_2\SPvalley_2}^{B^{\prime}}(\mathbf{r})][\psi_{\SPorbital\SPspin_3\SPvalley_3}^{B^{\prime}}(\mathbf{r})]^*[\psi_{\SPorbital\SPspin_4\SPvalley_4}^{B^{\prime}}(\mathbf{r})]>0$ (for all combinations of $\SPvalley_i$ corresponding to  inter- and intra-valley scattering channels induced by \eqn\eqref{eqn:HSR}), captures specific dot state characteristics, i.e.,  dot shape, gap, and mode number. The short-range interaction constants, $g_{ij}$, are a priori unknown and we discuss possible level orderings for different values of these couplings in \sect\ref{sec:TPStates}. In a finite magnetic field, the  \TP{}  levels split according to the g-factors in \eqn\eqref{eqn:sEnergies}. The valley g-factor, $\gVTPs$, of the \TP{} states computes as the sum of the \SP{} g-factors in the two valleys. For valley polarized states, $\gVTPs$  exceeds the \SP{} valley and spin g-factors. Conversely, the  g-factors from both valleys cancel for any valley coherent \TP{} state.


The ten possible \TP{} states with orbitally antisymmetric (\TPasymm)  wave function are,
\begin{align}
\nonumber \StatenmmminusOneTminusZ=&\dotc_{\SPorbital\downminus}\dotc_{\SPorbitalm\downminus}|0\rangle,\\
\nonumber \StatenmmZeroTminusZ=&\frac{1}{\sqrt{2}}(\dotc_{\SPorbital\upminus}\dotc_{\SPorbitalm\downminus}+\dotc_{\SPorbital\downminus}\dotc_{\SPorbitalm\upminus})|0\rangle,\\
\nonumber \StatenmmplusOneTminusZ=&\dotc_{\SPorbital\upminus}\dotc_{\SPorbitalm\upminus}|0\rangle,\\
\nonumber  \StatenmmminusOneTplusX=&\frac{1}{\sqrt{2}}(\dotc_{\SPorbital\downplus}\dotc_{\SPorbitalm\downminus}+\dotc_{\SPorbital\downminus}\dotc_{\SPorbitalm\downplus})|0\rangle,\\
\nonumber \StatenmmZeroTplusX=&\frac{1}{2}(\dotc_{\SPorbital\upplus}\dotc_{\SPorbitalm\downminus}+\dotc_{\SPorbital\upminus}\dotc_{\SPorbitalm\downplus}\\
\nonumber&+\dotc_{\SPorbital\downplus}\dotc_{\SPorbitalm\upminus}+\dotc_{\SPorbital\downminus}\dotc_{\SPorbitalm\upplus})|0\rangle,\\
\nonumber \StatenmmplusOneTplusX=&\frac{1}{\sqrt{2}}(\dotc_{\SPorbital\upplus}\dotc_{\SPorbitalm\upminus}+\dotc_{\SPorbital\upminus}\dotc_{\SPorbitalm\upplus})|0\rangle,\\
\nonumber \StatenmmminusOneTplusZ=&\dotc_{\SPorbital\downplus}\dotc_{\SPorbitalm\downplus}|0\rangle,\\
\nonumber \StatenmmZeroTplusZ=&\frac{1}{\sqrt{2}}(\dotc_{\SPorbital\upplus}\dotc_{\SPorbitalm\downplus}+\dotc_{\SPorbital\downplus}\dotc_{\SPorbitalm\upplus})|0\rangle,\\
\nonumber \StatenmmplusOneTplusZ=&\dotc_{\SPorbital\upplus}\dotc_{\SPorbitalm\upplus}|0\rangle,\\
\nonumber \StatenmSzeroTminusX =&\frac{1}{2}(\dotc_{\SPorbital\upplus}\dotc_{\SPorbitalm\downminus}-\dotc_{\SPorbital\upminus}\dotc_{\SPorbitalm\downplus}\\
&-\dotc_{\SPorbital\downplus}\dotc_{\SPorbitalm\upminus}+\dotc_{\SPorbital\downminus}\dotc_{\SPorbitalm\upplus})|0\rangle.
\label{eqn:asStates}
\end{align}
For brevity we consider  the simplest case where the \TP{} \GS{} consists of exactly two \SP{} orbitals $\SPorbital$ and $\SPorbitalm$ (substantial admixing of more than two orbitals is relevant only at higher energies \cite{knotheQuartetStatesTwoelectron2020}).
 The energies of the states in \eqn\eqref{eqn:asStates} are,
\begin{align}
\nonumber& E_{\LnmmminusOneTminusZ}\\
\nonumber&=\TPEn^{\TPasymm}_{ \SPorbital\SPorbitalm} -\gVTPa \mu_{B}B + \Ecap(2) +2\Delta_{SO}-g\mu_{B} B ,\\
\nonumber &E_{\LnmmZeroTminusZ}=\TPEn^{\TPasymm}_{ \SPorbital\SPorbitalm} + \Ecap(2) -\gVTPa \mu_{B}B ,\\
\nonumber& E_{\LnmmplusOneTminusZ}\\
\nonumber& =\TPEn^{\TPasymm}_{ \SPorbital\SPorbitalm}-\gVTPa \mu_{B}B + \Ecap(2) -2\Delta_{SO}+g\mu_{B} B ,\\
\nonumber &E_{\LnmmminusOneTplusX}=\TPEn^{\TPasymm}_{ \SPorbital\SPorbitalm} + \Ecap(2) -g\mu_{B} B  ,\\
\nonumber &E_{\LnmmZeroTplusX}=\TPEn^{\TPasymm}_{ \SPorbital\SPorbitalm} + \Ecap(2)  ,\\
\nonumber &E_{\LnmmplusOneTplusX}=\TPEn^{\TPasymm}_{\SPorbital\SPorbitalm}  + \Ecap(2) +g\mu_{B} B ,\\
\nonumber &E_{\LnmmminusOneTplusZ}\\
\nonumber &=\TPEn^{\TPasymm}_{ \SPorbital\SPorbitalm} +\gVTPa \mu_{B}B + \Ecap(2) -2\Delta_{SO}-g\mu_{B} B ,\\
\nonumber &E_{\LnmmZeroTplusZ}=\TPEn^{\TPasymm}_{ \SPorbital\SPorbitalm} +\gVTPa \mu_{B}B + \Ecap(2)  ,\\
\nonumber &E_{\LnmmplusOneTplusZ}\\
 \nonumber& =\TPEn^{\TPasymm}_{ \SPorbital\SPorbitalm}  +\gVTPa \mu_{B}B + \Ecap(2)+2\Delta_{SO} +g \mu_{B}B,\\
& E_{\LnmSzeroTminusX}=\mathfrak{E}^{\TPasymm}_{\SPorbital\SPorbitalm}  + \Ecap(2) .
 \label{eqn:asEnergies}
\end{align}
Here, {$\TPEn^{\TPasymm}_{ \SPorbital\SPorbitalm}$ is the energy of the orbitally antisymmetric states of two screened interacting electrons in \SP{} orbitals $\SPorbital$ and $\SPorbitalm$ (akin to the orbitally symmetric state described above),  
and   $\gVTPa$  is the  valley g-factor of the \TP{} multiplet.

\textbf{Coupling to the leads.} 
The point contacts in the \BLG{} channel to the left and right of the \QD{} provide discrete lead modes due to the transverse confinement. These modes can couple to the \QD{}. Close to pinching off the lowest of their modes, we can treat the \QPC{}s as tunnel junctions with tunnelling amplitudes $t^L$ ($t^R$) for the left (right) \QPC{}. We describe these single-channel leads with a Hamiltonian, 
\begin{equation}
\HLeads=\sum_{l=L,R}\sum_{k,\SPspin,\SPvalley}\epsilon^{l}_k\leadc_{lk\SPspin\SPvalley}\leada_{lk\SPspin\SPvalley},
\end{equation}
where $\leadc_{lk\SPspin\SPvalley}$ creates a lead electron with momentum k, energy $\epsilon^{l}_{k}$, spin $\SPspin$, and valley quantum number  $\SPvalley$.
The lead-dot tunnelling Hamiltonian is given by, 
\begin{equation}
\HTun=\sum_{l=L,R}\sum_{n,k,\SPspin,\SPvalley} \big(t^{l}_{\SPorbital\SPspin\SPvalley}\leadc_{lk\SPspin\SPvalley}\dota_{\SPorbital\SPspin\SPvalley} +\text{h.c.}\big).
\label{eqn:tun}
\end{equation}
In the following sections, we use this tunnelling Hamiltonian in \eqn\eqref{eqn:tun} to compute the tunnelling current across the \BLG{} \QD{}.

\section{Tunneling rates and rate equations}
\label{sec:Tunnel}
 For our transport calculations, we  consider the high-temperature regime 
 \begin{align}
\nonumber & \broad_{\SPorbital\SPspin\SPvalley}\ll k_{B}T<| g_{\perp,zz}\intW|<\Delta E_{\partnumb,\partnumb\pm 1},\\
 & \broad_{\SPorbital\SPspin\SPvalley}=\pi\sum_{l=L,R}\nu_{l}|t_{\SPorbital\SPspin\SPvalley}^{l}|^{2},
 \end{align}
  where in the  last term  we compare to the energy difference of dot states with different particle number and 
  $\broad_{\SPorbital\SPspin\SPvalley}$ is the tunnel-coupling induced level broadening with lead density of states $\nu_{l}$. 
  For a level broadening $\broad_{\SPorbital\SPspin\SPvalley}$ much smaller than the thermal energy $k_{B}T$, we can compute transport perturbatively in the tunnel Hamiltonian\cite{golovachTransportDoubleQuantum2004, aleinerQuantumEffectsCoulomb2002, begemannInelasticCotunnelingQuantum2010}, $\HTun$, in \eqn\eqref{eqn:tun}.

The lowest (first) order in the lead-dot tunnel coupling describes the single-electron processes involved in sequential tunnelling: an electron tunnels either from the leads to the dot or from the dot to the leads thereby changing the occupation number of the dot by one. Expanding to first order in $\HTun$ and applying Fermi's golden rule, transition rates  for a one-electron tunnelling, which induces a transition  of dot from a \SP{} state,  $|\partnumb^{\prime}=1:\chi^{\prime}\rangle$, to  a \TP{} dot state,  $|\partnumb=2:\chi\rangle$, read,
\begin{widetext}
\begin{equation}
W_{2:\chi\leftarrow1:\chi^{\prime}}=\frac{2\pi}{\hbar}\sum_{f,i}| \LD f |\LD 2:\chi |\HTun | 1:\chi^{\prime}\RD| i\RD|^2  \,\rho^{i}\,\delta(E_{f,2:\chi}-E_{i,1:\chi^{\prime}}) =\!\!\! \sum_{{l=L,R}}\frac{2\pi}{\hbar}|t^{l}_{\SPorbital\SPspin\SPvalley}|^{2}f(E_{2:\chi}-E_{1:\chi^{\prime}}-\Chempot^{l})=\!\!\! \sum_{{l=L,R}}W^{l}_{2:\chi\leftarrow1:\chi^{\prime}}.
\label{eqn:WSeqT}
\end{equation}
\end{widetext}
Here, $\partnumb$ indicates the dot particle number  and $\chi$ identifies the state of the corresponding multiplet. Hence, a prefix $\partnumb=1$ implies $\chi=(\SPorbital,\SPspin,\SPvalley)$ and for $\partnumb=2$, $\chi$ indexes the orbital, spin, and valley combinations from the family of states in \eqns\eqref{eqn:sStates} or \eqref{eqn:asStates}, respectively. In \eqn\eqref{eqn:WSeqT}, $(\SPorbital\SPspin\SPvalley)$ are the indices of the electron tunnelling into the dot,  forming the \TP{} state $|2\!:\chi\rangle$  with the single electron previously on the dot (the latter having  quantum numbers $\chi^{\prime}$). The initial and final states of the leads are $|i\rangle=|i_{L}\rangle|i_{R}\rangle$ and $|f\rangle=|f_{L}\rangle|f_{R}\rangle$, the former weighted by a thermal distribution $\rho^{i}$. Further, $f$ denotes the Fermi function and $\Chempot^{l}$ is the chemical potential of lead $l$, which depends on the bias voltage, $\Vbias$. We consider the case where the dot is biased symmetrically, $\Chempot^{L/R}=\pm e \Vbias$, with respect to the equilibrium chemical potential.
The rates for the reverse transitions, $|1\!:\chi^{\prime}\rangle\leftarrow|2\!:\chi\rangle$,  follow from \eqn\eqref{eqn:WSeqT} by replacing $f(E) \rightarrow 1 - f(E)$. We provide the explicit rates for each transition in \app\ref{app:SeqTunRates}. 

Going to second order in $\HTun$ describes correlated two-electron cotunnelling: an electron tunnels from one lead to the other (or the same lead) via the \QD{}, leaving the occupation number of the dot invariant. Within each particle number sector ($\partnumb=1$ or $\partnumb=2$), the dot's state may change (inelastic cotunnelling) or remain the same (elastic cotunnelling).
The corresponding cotunnelling rates read,
\begin{widetext}
\begin{align}
 &\nonumber W_{1:\chi\leftarrow1:\chi^{\prime}}=\sum_{l,\l^{\prime}}   W^{l,\l^{\prime}}_{1:\chi\leftarrow1:\chi^{\prime}}\\
 &\nonumber=\frac{2\pi}{\hbar}\sum_{l,\l^{\prime},\tilde\chi}|t_{\chi^{\prime}}^{l}\; t_{\chi}^{l^{\prime}}\hskip0pt^*|^2 \iint d\epsilon_{k}^{l^{\prime}}d\epsilon_{k^{\prime}}^{l}\Big| \frac{1}{E_{i,1:\chi^{\prime}}-E_{2:{\tilde{\chi}}}+\epsilon_{k}^{l^{\prime}}+i0^{+}} \Big|^{2} 
  f(\epsilon_{k}^{l^{\prime}}-\Chempot^{l^{\prime}} ) \big[ 1-f(\epsilon_{k^{\prime}}^{l}-\Chempot^{l} ) \big] \delta (E_{1:\chi} + \epsilon_{k^{\prime}}^{l}-E_{1:\chi^{\prime}}-\epsilon_{k}^{l^{\prime}}),\\
   &\nonumber W_{2:\chi\leftarrow2:\chi^{\prime}}=\sum_{l,\l^{\prime}}   W^{l,\l^{\prime}}_{2:\chi\leftarrow2:\chi^{\prime}}\\
 &=\frac{2\pi}{\hbar}\sum_{l,\l^{\prime},\tilde\chi}|t_{\chi^{\prime}}^{l}\; t_{\chi}^{l^{\prime}}\hskip0pt^*|^2 \iint d\epsilon_{k}^{l^{\prime}}d\epsilon_{k^{\prime}}^{l}\Big| \frac{1}{E_{i,2:\chi^{\prime}}-E_{1:{\tilde{\chi}}}-\epsilon_{k}^{l^{\prime}}+i0^{+}} \Big|^{2} 
  f(\epsilon_{k}^{l^{\prime}}-\Chempot^{l^{\prime}} ) \big[ 1-f(\epsilon_{k^{\prime}}^{l}-\Chempot^{l} ) \big] \delta (E_{2:\chi} + \epsilon_{k^{\prime}}^{l}-E_{2:\chi^{\prime}}-\epsilon_{k}^{l^{\prime}}).
 \label{eqn:WCoT}
\end{align}
\end{widetext}
These rates involve the intermediate states of higher or lower dot occupation number, $\partnumb\pm1$, if they are allowed by spin and valley selection rules. In \eqn\eqref{eqn:WCoT}, we take into account transitions via the \SP{} \GS{} multiplet and the \TP{} \GS{} multiplets of \eqn\eqref{eqn:sStates} and \eqref{eqn:asStates}. Projection onto these \SP{} and \TP{} state spaces is  valid for \QD{}s where all other states  are separated sufficiently  in energy to exclude any virtual transitions to them. It is not straightforward to evaluate the cotunneling rates in \eqn\eqref{eqn:WCoT} due to the second-order poles causing the integrals to diverge. These divergences are related to the intermediate state's zero width and hence infinite lifetime within this perturbative approach. We follow the standard regularization procedure to extract the correct cotunneling rates from \eqn\eqref{eqn:WCoT} \cite{begemannInelasticCotunnelingQuantum2010, kochTheoryFranckCondonBlockade2006, kochThermopowerSinglemoleculeDevices2004, turekCotunnelingThermopowerSingle2002, averinPeriodicConductanceOscillations1994}: First, a level width $\gamma \sim \broad_{\SPorbital\SPspin\SPvalley}$ is introduced as imaginary parts in the denominators (accounting for the intermediate states' tunnel-coupling induced level broadening). These imaginary parts shift the poles away from the real axis, and the integrals can be carried out. Next, the resulting expression is expanded in powers of $\gamma$. The leading order term is 
a sequential-tunnelling contribution (reflecting that, at finite temperature, the final state of any cotunneling-induced transition can also be reached via two successive single-electron tunnelings). This term is disregarded to avoid double-counting sequential tunnelling processes. The next-to-leading-order term in the $\gamma$ expansion gives the regularized expression for the cotunneling rate, where the limit $\gamma\rightarrow0$ can be taken. We provide the regularization calculations and resulting expressions for the cotunnelling rates $W_{\partnumbcol\chi\leftarrow\partnumbcol\chi^{\prime}}$ in appendix \ref{app:CoTunRates}.

Given the rates for transitions between different dot states, we write a master equation describing the dynamics of the probabilities, $\dotprob_{\dotstate}$, for the state, $|\dotstate\rangle$, to be occupied at a given time,
\begin{equation}
\dot{\dotprob}_{\dotstate}=\sum_{\dotstateP}(W_{\dotstate\leftarrow \dotstateP}\dotprob_{\dotstateP}-W_{\dotstateP\leftarrow \dotstate}\dotprob_{\dotstate}),
\label{eqn:rateeqn}
\end{equation}
where the terms with changing particle number, $1\leftrightarrows 2$, describe current flow whereas cotunnelling terms introduce relaxation within the multiplets at fixed particle number. 
We solve these rate equations, \eqn\eqref{eqn:rateeqn}, in the stationary limit, $\dot\dotprob_{\dotstate}=0$, using the normalization condition $\sum_{\dotstate}\dotprob_{\dotstate}=1$.
From the probabilities we compute the total particle current $I= I_{seq}+ I_{cot}$, with the sequential tunnel currents flowing from the dot to lead $l$,
\begin{equation} 
 I_{seq}^{l}=\sum_{1:\chi,2:\chi^{\prime}} (W^l_{1:\chi \leftarrow 2:\chi^{\prime}})e\dotprob_{2:\chi^{\prime}}-(W^l_{2:\chi^{\prime}\leftarrow1:\chi})e\dotprob_{1:\chi},
\end{equation}
and the cotunneling current between lead $l^{\prime}$ and $l$,
\begin{equation}
 I_{cot}^{l}=\sum_{\dotstate,\dotstate^{\prime}} (W^{ll^{\prime}}_{\dotstate^{\prime}\leftarrow\dotstate}-W^{l^{\prime}l}_{\dotstate^{\prime}\leftarrow\dotstate})e\dotprob_{\dotstate}.
\end{equation}


It depends on the tunnelling strength compared to the isolated dot's level splitting whether second-order cotunnelling processes contribute significantly to transport. We define the regime of \emph{purely sequential tunnelling} for weak dot-lead tunnel coupling, and the regime of \emph{sequential + cotunnelling} for stronger dot-lead tunnel coupling, where second order effects contribute. Numerically, we find that the first regime is realized for $|t^{l}_{\SPorbital\SPspin\SPvalley}|\sim\Delta E_{\partnumb,\partnumb\pm1}/1000$ while reaching the latter regime requires approximately $|t^{l}_{\SPorbital\SPspin\SPvalley}|\sim\Delta E_{\partnumb,\partnumb\pm1}/100$.


\section{Resolving the \TP{} dot states}
\label{sec:TPStates}

 \begin{figure*}[htb!]
 \centering
\includegraphics[width=0.8\linewidth]{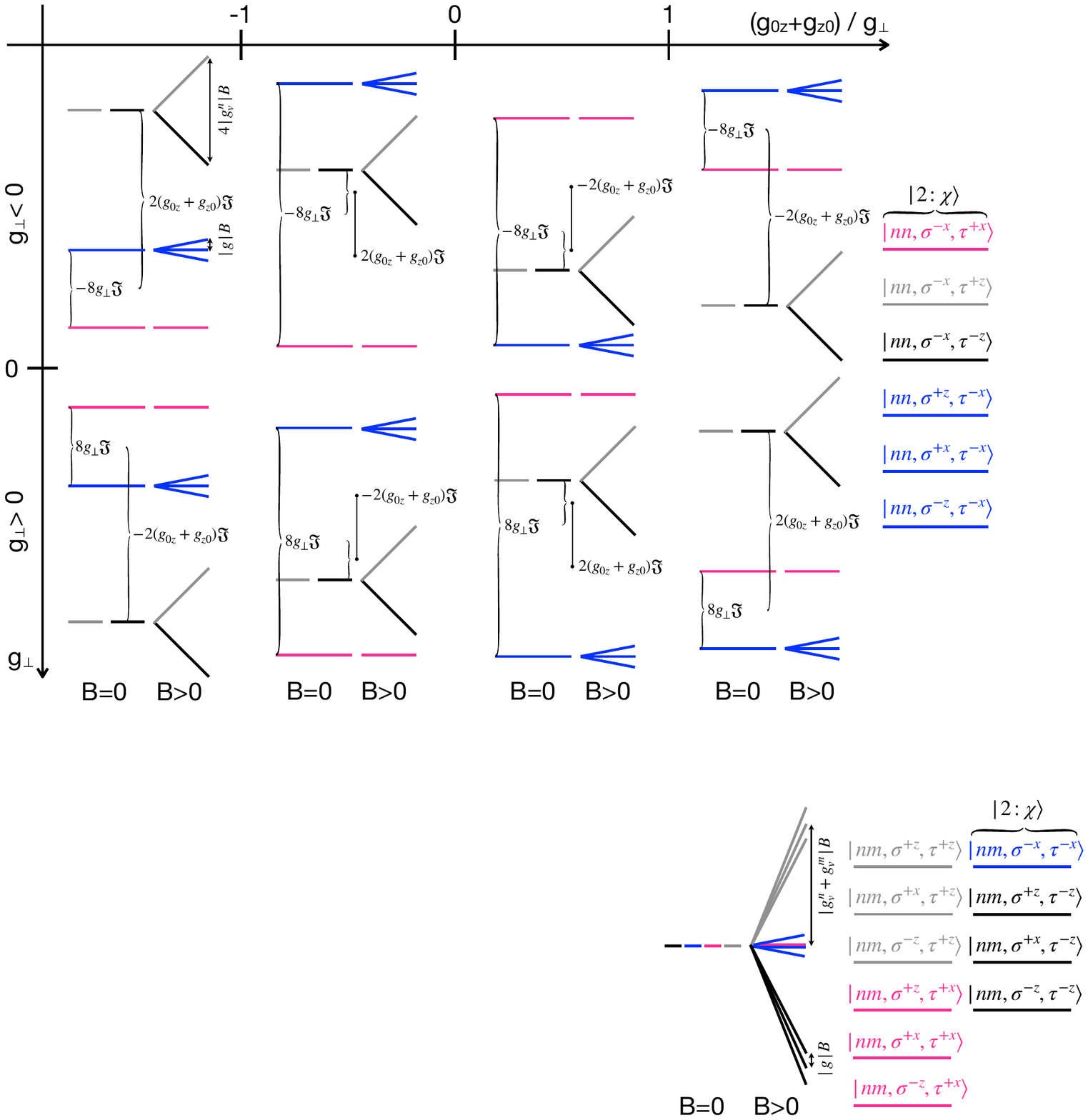}
\caption{The ordering of the  dot's \TP{} lowest-state multiplet with symmetric orbital wave function at zero bias voltage, \eqn\eqref{eqn:sEnergies}, depend on the relative magnitude and sign of the short-range interaction coupling constants, $g_{\perp},\; g_{0z}$, and $g_{z0}$, and on the magnetic field.}
\label{fig:SSplittings}
\end{figure*}

\subsection{Spectroscopy of an orbitally symmetric \TP{} ground state}
\label{sec:S}

This section considers dots with orbitally symmetric \TP{} ground state wave functions.  We discuss the  possible level orderings which can result from \eqn\eqref{eqn:sEnergies} and at zero and finite magnetic field and how to distinguish the spin and valley states in tunnelling transport.

\textbf{Possible level orderings of orbitally symmetric \TP{} dot states.}
We illustrate the various level orderings of the states in \eqn\eqref{eqn:sStates} for different signs and relative magnitudes of the short-range couplings $g_{0z}, g_{z0},$ and $g_{\perp}$ in \fig\ref{fig:SSplittings}. Generally, there are three levels at zero magnetic field, being singly, doubly, and three-fold degenerate, respectively. These degeneracies are lifted by a finite magnetic field, splitting different valley and spin states. According to \eqn\eqref{eqn:sEnergies}, the coupling constant $g_{zz}$ shifts all energies equally.  The mutual splitting between the two inter-valley coherent states, $\tau^{\pm x}$, is proportional to the coupling $g_{\perp}$,  while these states are split from the valley polarized states, $\tau^{\pm z}$, proportionally to the sum $g_{0z} + g_{z0}$.

Tunnelling transitions in the single- and \TP{} sector allow identifying the spin and valley states and determining the short-range couplings by combining the two following considerations: Firstly, the single- and \TP{} states split in a magnetic field.  
 Transition energies   hence depend on the difference in  single- and \TP{} valley g-factors. Besides, any \SP{}-to-\TP{} tunnelling transition is subject to spin and valley selection rules. Therefore, we can identify the \TP{} states that can be reached, e.g., from the \SP{} \GS{}.
With the \TP{} levels being identified, we can relate the level splittings to the short-range interaction couplings $g_{0z}, g_{z0},$ and $g_{\perp}$ as in \fig\ref{fig:SSplittings}. Hence, classifying the dot's \TP{} states and their mutual gaps is a way to quantify \BLG{}'s microscopic short-range interaction parameters.  

\begin{figure*}[t!]
 \centering
\includegraphics[width=0.8\linewidth]{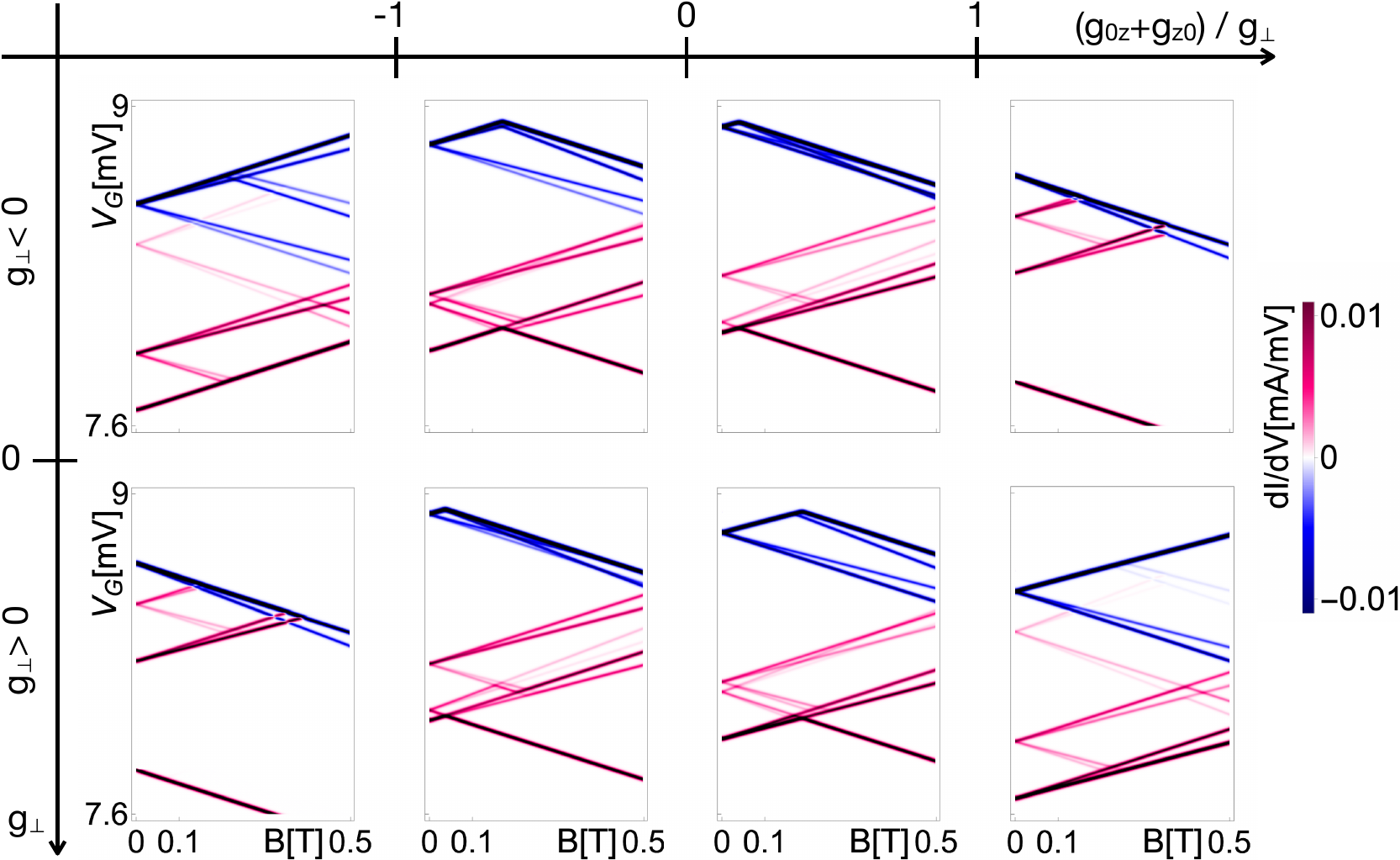}
\caption{{Differential conductance maps for the \TP{} multiplets in \fig\ref{fig:SSplittings} for the different possible regimes of short-range interaction constants $g_{\perp}$, $g_{0z}$, and $g_{z0}$.  (sequential tunnelling and symmetric coupling to both leads). The conductance increases/decreases when a \SP{}-to-\TP{} transition enters/leaves the bias window at fixed bias voltage $V_{B}=0.45$ mV (red/blue lines).  Permissible single-to-\TP{} transitions depend on spin and valley selection rules and whether the \TP{} \GS{} is spin and valley coherent or polarized.  The difference of the \TP{} and \SP{} valley g-factors dominates the slope of the lines with magnetic field. Here, $k_{B}T$= 0.003 meV.}}
\label{fig:SAll}
\end{figure*}



 \subsubsection{Two-particle states with broken time-inversion symmetry}
\label{sec:S-TRSB}
Single-to-\TP{} transitions to the levels in \fig\ref{fig:SSplittings} yield differential conductance features as in \fig\ref{fig:SAll}. Here, we consider sequential tunnelling  and symmetric coupling to the leads. Differential conductance maps as the ones in \fig\ref{fig:SAll} are cuts at finite bias voltage (we chose $V_{B}=0.45$ meV) through the Coulomb diamonds for different values of magnetic field (cf.~\fig\ref{fig:Teaser}). Each allowed single-to-\TP{} transition manifests as an increase/decrease in conductance (red/blue lines) once this transition enters/leaves the bias window. The differential conductance features at zero magnetic field reflect the splittings of the \TP{} multiplets  in \fig\ref{fig:SSplittings}. At finite magnetic field, the conductance lines disperse according to the \TP{} and \SP{} g-factors. Hence, for similar zero-field splittings and similar g-factors, the conductance maps can coincide even for distinct \TP{} level orderings.

\begin{figure*}[t!]
 \centering
\includegraphics[width=1\linewidth]{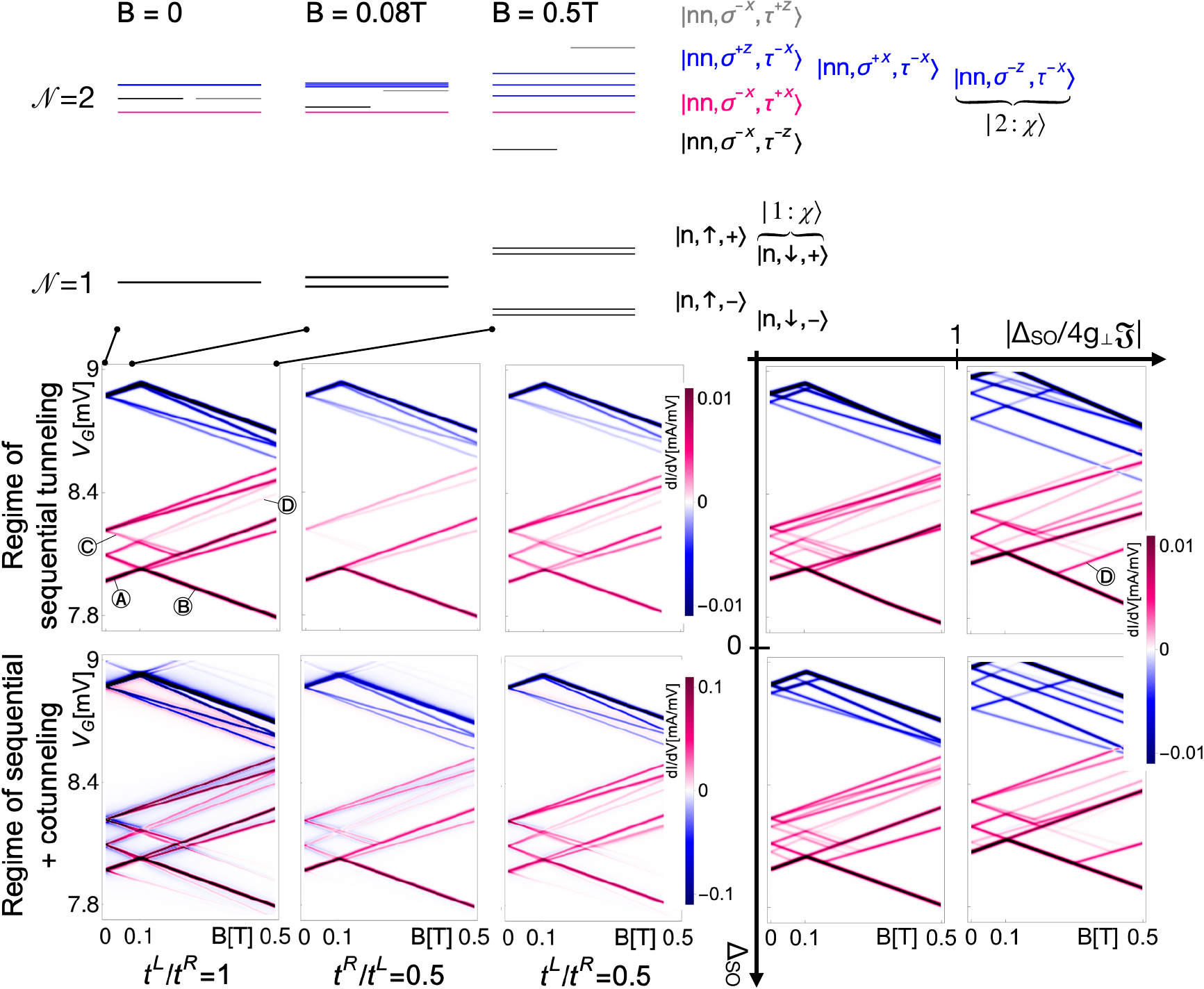}
\caption{{Transport characteristics of a \BLG{} \QD{} with an orbitally symmetric \TP{} \GS{} preserving time-inversion invariance (in which case, $g_{z0}=g_{0z}=0$, compared to \fig\ref{fig:SAll}). Top: Single-particle and \TP{} dot levels for different magnetic fields. Bottom left: Differential conductance across the dot in different tunnelling regimes and symmetric or asymmetric lead coupling (suppressing coupling to the right lead = source or the left lead = drain, respectively) depending on a magnetic field, $B$. For the conductance maps, we fix $V_{B}=0.45$ mV and $k_{B}T$= 0.003 meV. The sequential tunnelling regime is realized for $|t^{l}_{\SPorbital\SPspin\SPvalley}|\sim\Delta E_{\partnumb,\partnumb\pm1}/1000$ while significant cotunnelling contributions  require approximately $|t^{l}_{\SPorbital\SPspin\SPvalley}|\sim\Delta E_{\partnumb,\partnumb\pm1}/100$. Cotunnelling induces relaxation processes within each fixed particle number multiplet and hence opens additional transport channels compared to purely sequential tunnelling. The panels on the right consider the potential influence of a finite \SO{} coupling gap, $\Delta_{SO}$, possibly smaller (here $|\Delta_{SO}|=0.02$ meV) or larger (here $|\Delta_{SO}|=0.1$ meV) than the splitting induced by $g_{\perp}$ and of different sign.} The labels \raisebox{.5pt}{\textcircled{\raisebox{-1pt} {{\fontfamily{phv}\selectfont A}}}}, \raisebox{.5pt}{\textcircled{\raisebox{-1pt} {{\fontfamily{phv}\selectfont B}}}}, \raisebox{.5pt}{\textcircled{\raisebox{-1pt} {{\fontfamily{phv}\selectfont C}}}}, \raisebox{.5pt}{\textcircled{\raisebox{-1pt} {{\fontfamily{phv}\selectfont D}}}} indicate transitions which we discuss in detail in the main text.}
\label{fig:WeakMain}
\end{figure*}

To facilitate the electron transport through a dot, the bias window must allow   \SP{}-to-\TP{} transitions between the \GS{} of the dot with one and two electrons, respectively.
 When the \TP{} \GS{} is valley coherent, the corresponding lines in the differential conductance maps have positive slope in a magnetic field. While these coherent \TP{} states do not disperse with $B$, the \SP{} \GS{}, $|\SPorbital,\downarrow, -\rangle$ is pushed down and the energy required for this transition increases. Conversely, a $K^{-}$ valley polarized \TP{} \GS{}s is pushed down even faster with $B$ (since  $\gVTPs >\gVSP$), causing the transitions energy to decrease. This leads to lines with negative slopes limiting the bias window range in \fig\ref{fig:SAll} for these cases. Within the bias window, whether energetically allowed \SP{}-to-\TP{} transitions contribute to transport is determined by spin and valley selection rules. For example, the $K^{+}$ excited \SP{} states can be populated via transitions to valley coherent \TP{} states. However, if there are no such transitions available at equal or lower gate voltage,  the $K^{+}$ \SP{} states are depopulated, causing the corresponding lines to terminate in the differential conductance maps in \fig\ref{fig:SAll}.

\subsubsection{Two-particle states preserving time-inversion invariance}
\label{sec:S-TRS}
In the following sections \ref{sec:S-TRS}, \ref{sec:AS} and \ref{sec:Cross}, we exemplify the \QD{}'s tunnelling characteristics for one specific level arrangement  of the orbitally symmetric \TP{} states and study different regimes of lead couplings as well as the impact of a finite \SO{} coupling gap. Numerical values we have  estimated  previously in one specific dot model\cite{knotheQuartetStatesTwoelectron2020}, yielded  $g_{0z}=g_{z0}=0,$ (preserving time-reversal invariance),  $\;g_{zz}>0, \;g_{\perp}<0$ (favouring the spin and valley coherent \GS{} $\StatenSzeroTminusZ$), and $\intW g_{zz}\gg4|\intW g_{\perp}|$. For this choice of short-range couplings, the \TP{} triplet is equally spaced at $B=0$ (top left panel of  \fig\ref{fig:WeakMain}).  A finite magnetic field splits the levels according to the spin and valley configuration (top row of  \fig\ref{fig:WeakMain}).

 \begin{figure}[t!]
 \centering
\includegraphics[width=1\linewidth]{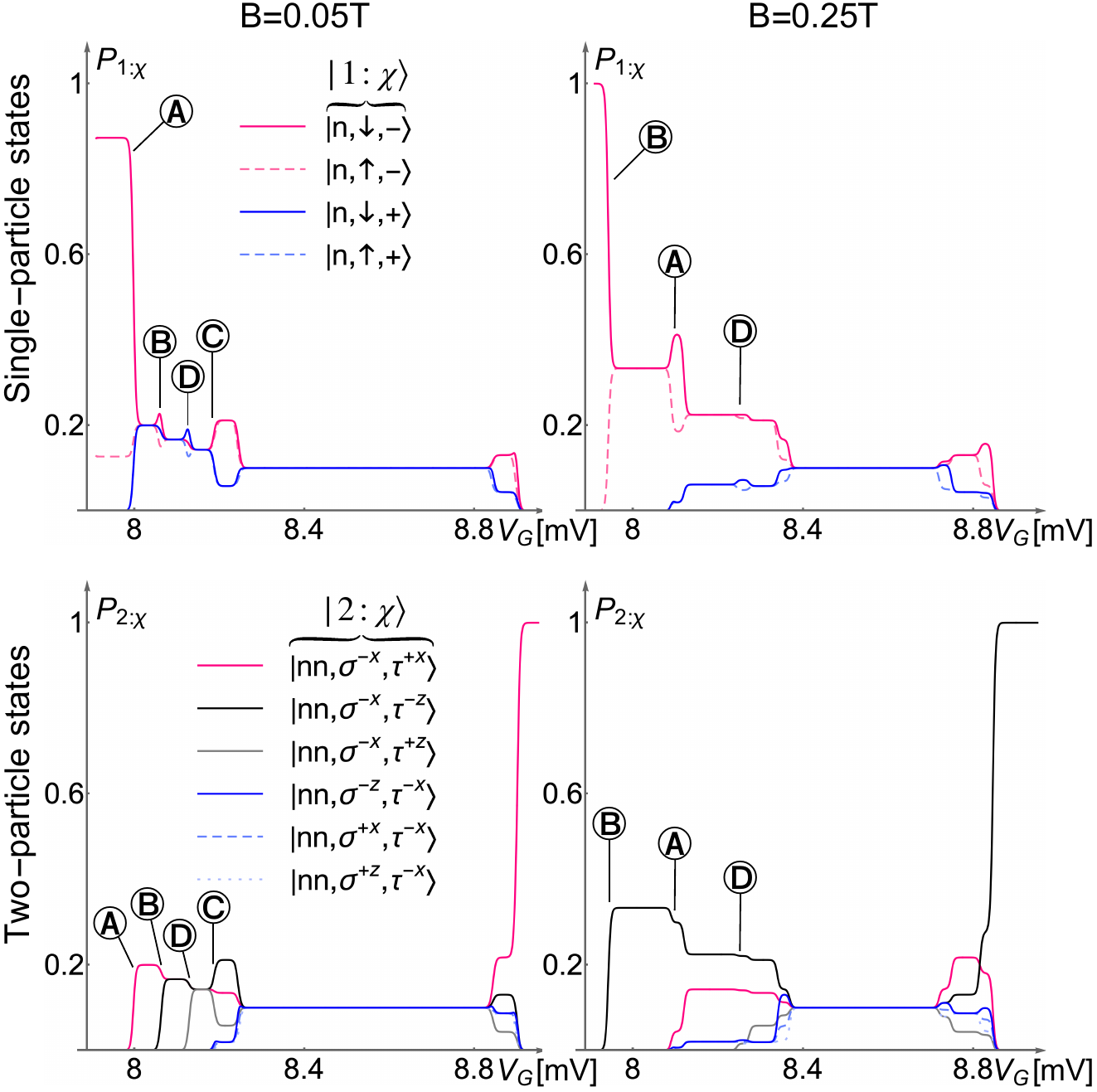}
\caption{{ Probabilities for the \SP{} and  orbitally symmetric \TP{} states to be occupied for fixed magnetic field cuts  along the gate voltage axis through the symmetrically coupled differential conductance maps with zero \SO{} coupling in \fig\ref{fig:WeakMain} (leftmost differential conductance maps) in the sequential tunnelling regime.}}
\label{fig:WeakPop}
\end{figure}

The contrasting magnetic field coupling of valley polarized and valley coherent \TP{} states leads to level crossings at finite $B$. For zero and small magnetic field the state $\StatenSzeroTplusX$ is the \TP{} \GS{}. This spin- and valley-coherent state does not couple to the magnetic field. At sufficiently large $B$, the valley polarized state, $\StatenSzeroTminusZ$, being pushed down by the magnetic field, becomes the \TP{} \GS{}.
Being able to identify the differential conductance characteristics in \fig\ref{fig:WeakMain} with the possible \SP{}-to-\TP{} transitions allows extracting information about a symmetric \TP{} dot state.

In the regime of sequential tunnelling, transport is possible, once the gate voltage sufficed to induce the \GS{}-to-\GS{} transition. For zero or weak magnetic field, this is the transition,
\begin{equation}
  \raisebox{.5pt}{\textcircled{\raisebox{-1pt} {{\fontfamily{phv}\selectfont A}}}}: \;|\SPorbital,\downarrow,-\rangle\rightarrow\StatenSzeroTplusX.
   \end{equation}
 The involved \TP{} state occupies all four different spin and valley states. Hence, when one electron leaves the dot in the subsequent tunnelling process, the remaining electron can be in any of the \SP{} states. Consequently, with increasing gate voltage, all \SP{}-to-\TP{} transitions become possible and manifest in the differential conductance maps within the bias window. At higher magnetic fields, the \GS{}-to-\GS{} transition changes to,
   \begin{equation}
  \raisebox{.5pt}{\textcircled{\raisebox{-1pt} {{\fontfamily{phv}\selectfont B}}}}:\;     |\SPorbital,\downarrow,-\rangle\rightarrow\StatenSzeroTminusZ,
  \end{equation} 
    where the valley $K^{-}$ polarized \TP{} state entails that after the next tunnelling process, the remaining electron occupies one of the $K^{-}$ \SP{} states. As a consequence, transitions from the $K^{+}$ \SP{} states do not contribute to conductance in this regime if there is no transition to a valley-coherent \TP{} state lower in gate voltage. The corresponding lines \raisebox{.5pt}{\textcircled{\raisebox{-1pt} {{\fontfamily{phv}\selectfont C}}}} terminate in the differential conductance maps in \fig\ref{fig:WeakMain}. We note that coupling stronger to the source (left lead in our convention) and suppressing the drain coupling (right lead) suppresses transport features from the transitions involving valley polarized \TP{} states. In comparison, stronger coupling to the drain decreases the amplitudes of all transport channels.
 A finite \SO{} coupling gap, $\Delta_{SO}$, further splits the states and corresponding transitions depending on its sign and magnitude relative to the short-range splittings. When the \SO{} gap overcomes the short-range couplings, $|\Delta_{SO}|>4| g_{\perp}\intW|$, transitions may occur in a different order, exemplified by the transition \raisebox{.5pt}{\textcircled{\raisebox{-1pt} {{\fontfamily{phv}\selectfont D}}}} in  \fig\ref{fig:WeakMain}. We depict representative differential conductance maps in the regime of sequential  tunnelling and symmetric lead-coupling in the bottom right panels of \fig\ref{fig:WeakMain}.

Cotunnelling  leads to relaxation processes within the multiplets of each seperate particle number sector and can hence make additional transport channels available. In the coupling regime where cotunnelling processes play a significant role, the transitions from the $K^{+}$  \SP{} states to the $\tau^{-x}$ \TP{} states (\raisebox{.5pt}{\textcircled{\raisebox{-1pt} {{\fontfamily{phv}\selectfont C}}}} in \fig\ref{fig:WeakMain}) reappear compared to the regime of purely sequential tunnelling as a result of population of these \SP{} states via the $|\SPorbital,\downarrow, -\rangle\rightarrow |\SPorbital,\uparrow,+\rangle,\; |\SPorbital,\downarrow,+\rangle$ cotunnelling transitions (cotunnelling assisted sequential tunnelling \cite{golovachTransportDoubleQuantum2004}). Besides, we observe features outside the Coulomb diamond, where cotunnelling events populate states that do not yet fall into the bias window for a certain gate voltage value. Increasing magnetic field and any asymmetry in the lead couplings suppress cotunneling-induced transport features. The former is due to energy differences between states growing with $B$, suppressing inelastic events. The latter suppression comes from the fact that at finite bias, the relevant contributions to cotunnelling scattering rates involve tunnelling at both leads (cf.~\eqn\eqref{eqn:WCoT}).

The occupation probabilities, $\dotprob_{\dotstate}$ shown in \fig\ref{fig:WeakPop} for different values of magnetic field support the conclusions above about states contributing to transport in different regimes. Allowed transitions manifest as steps where state occupation numbers change. Cotunnelling processes alter these steps by introducing alternative transitions between dot states (see appendix \ref{app:CoTunProps}).

 \begin{figure}[t!]
 \centering
\includegraphics[width=0.72\linewidth]{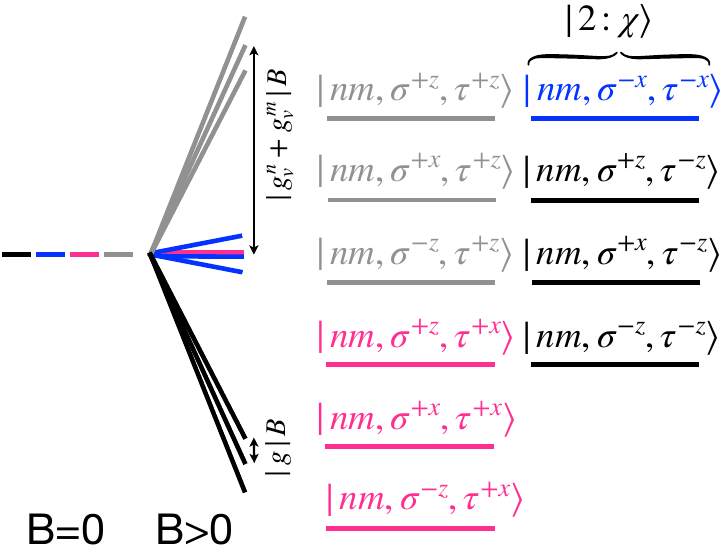}
\caption{Two-particle dot states with antisymmetric orbital wave function are degenerate at zero magnetic field and split linearly with $B$, cf.~ \eqn\eqref{eqn:asEnergies}. }
\label{fig:ASSplittings}
\end{figure}
\subsection{Spectroscopy of an orbitally antisymmetric \TP{} ground state}
\label{sec:AS}
\textbf{Level ordering of orbitally antisymmetric \TP{} dot states.}
\begin{figure}[t!]
 \centering
\includegraphics[width=1\linewidth]{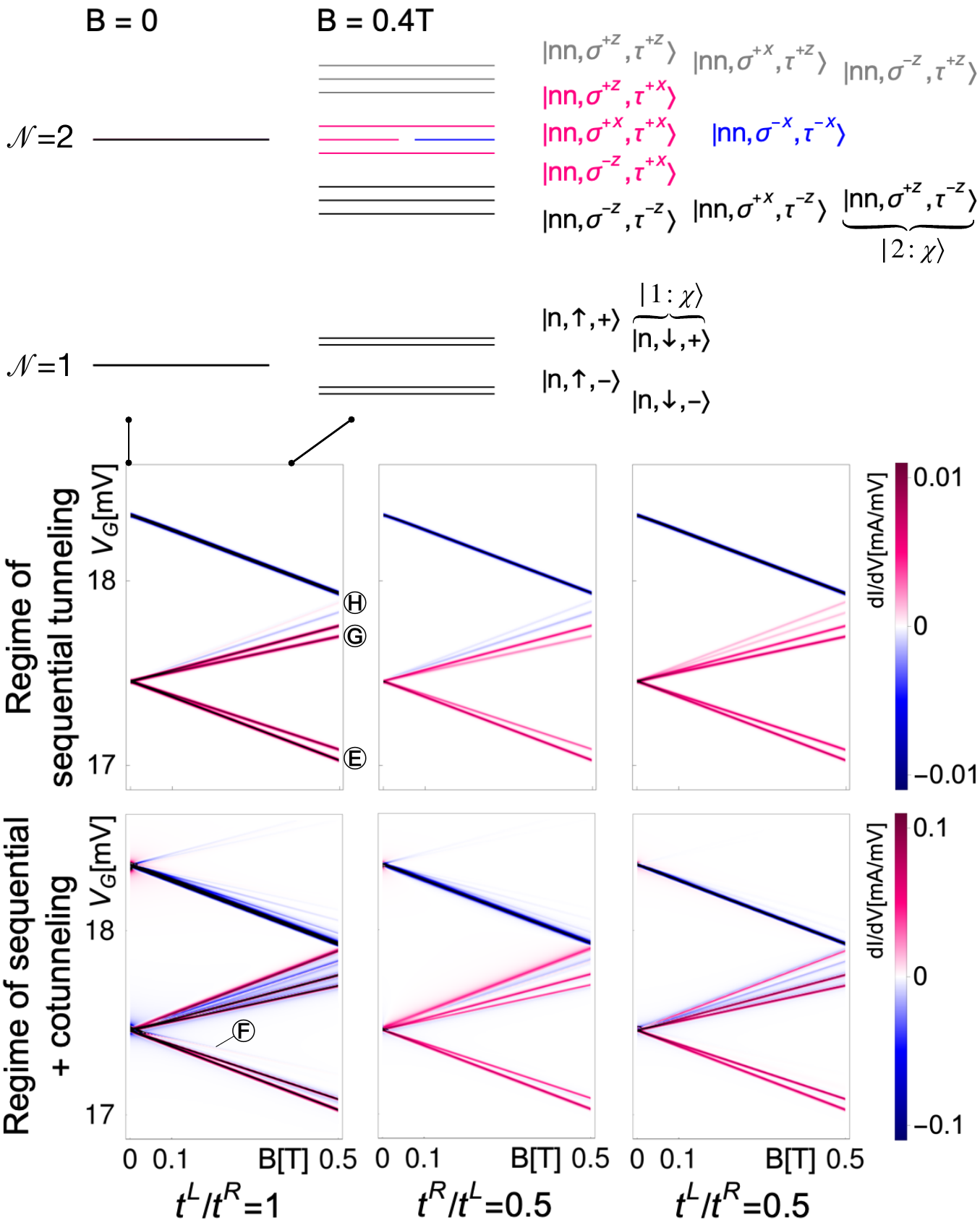}
\caption{{Level orderings and transport characteristics of a \BLG{} \QD{} in which two electrons form an orbitally antisymmetric wave function (parameters for the transport calculations as in \fig\ref{fig:WeakMain}). As opposed to an orbitally symmetric dot state (\fig\ref{fig:WeakMain}), there are no zero-field splittings and no inversion of the \TP{} multiplet's ordering, cf.~\fig\ref{fig:ASSplittings}. Note that here, we chose  $\Delta_{SO}=0$, a finite \SO{} gap yields two split copies of fans in the differential conductance maps.} The labels  \raisebox{.5pt}{\textcircled{\raisebox{-1pt} {{\fontfamily{phv}\selectfont E}}}},  \raisebox{.5pt}{\textcircled{\raisebox{-1pt} {{\fontfamily{phv}\selectfont F}}}},  \raisebox{.5pt}{\textcircled{\raisebox{-1pt} {{\fontfamily{phv}\selectfont G}}}},  \raisebox{.5pt}{\textcircled{\raisebox{-1pt} {{\fontfamily{phv}\selectfont H}}}} mark the transitions in \eqns\eqref{eqn:trans1} and \eqref{eqn:AddTrans}.}
\label{fig:StrongMain}
\end{figure}
The ten-fold degenerate spin and valley \TP{} states' multiplet  with orbitally antisymmetric wave function, \eqn\eqref{eqn:asStates}, splits in a magnetic field according to the states' spin and valley g-factors as in \fig\ref{fig:ASSplittings}. Hence, since $\gVSP\gg g$, identifying the allowed \SP{}-to-\TP{} transitions for tunnelling transport at finite $B$ yields information about the orbitally antisymmetric \TP{} dot state.
For non-zero magnetic field, $\StatenmmminusOneTminusZ$ is the \TP{} \GS{} (cf.~\fig\ref{fig:StrongMain} top row).
By purely sequential tunnelling, the following transitions are accessible,
\begin{align}
\nonumber \raisebox{.5pt}{\textcircled{\raisebox{-1pt} {{\fontfamily{phv}\selectfont E}}}}:\; &|\SPorbital,\downarrow,-\rangle, \; |\SPorbital,\uparrow,-\rangle\\\nonumber&\rightarrow \StatenmmminusOneTminusZ, \StatenmmZeroTminusZ, \StatenmmplusOneTminusZ, \\
\nonumber \raisebox{.5pt}{\textcircled{\raisebox{-1pt} {{\fontfamily{phv}\selectfont G}}}}:\;  &|\SPorbital,\downarrow,-\rangle, \; |\SPorbital\upminus\rangle\\\nonumber&\rightarrow  \StatenmmminusOneTplusX, \StatenmmZeroTplusX, \StatenmmplusOneTplusX ,\\
\nonumber&\hskip16pt \StatenmSzeroTminusX,\\
\nonumber \raisebox{.5pt}{\textcircled{\raisebox{-1pt} {{\fontfamily{phv}\selectfont H}}}}:\;  &|\SPorbital,\downarrow,+\rangle, \;|\SPorbital,\uparrow,+\rangle\\&\rightarrow \StatenmmminusOneTplusZ, \StatenmmZeroTplusZ, \StatenmmplusOneTplusZ.
\label{eqn:trans1}
\end{align}
Each transition in \eqn\eqref{eqn:trans1} contributes a line to the differential conductance maps in \fig\ref{fig:StrongMain}, the slopes of which are given by the g-factor difference of the involved \SP{} and \TP{} states.
The first transition listed in \eqn\eqref{eqn:trans1} is the \GS{}-to \GS{} transition. The  \SP{} excited state $|\SPorbital,\uparrow,-\rangle$ is populated from the spin  coherent \TP{} state $\StatenmmZeroTminusZ$ via the tunnelling sequence  $|\SPorbital,\downarrow,-\rangle\rightarrow\StatenmmZeroTminusZ\rightarrow |\SPorbital,\uparrow,-\rangle$. The transitions 
\begin{align}
\nonumber \raisebox{.5pt}{\textcircled{\raisebox{-1pt} {{\fontfamily{phv}\selectfont F}}}}:\;  &|\SPorbital,\downarrow,+\rangle,\;|\SPorbital,\uparrow,+\rangle\\\nonumber&\rightarrow  \StatenmmminusOneTplusX, \StatenmmZeroTplusX, \StatenmmplusOneTplusX ,\\
&\hskip16pt \StatenmSzeroTminusX,
\label{eqn:AddTrans}
\end{align}
 \begin{figure}[t!]
 \centering
\includegraphics[width=1\linewidth]{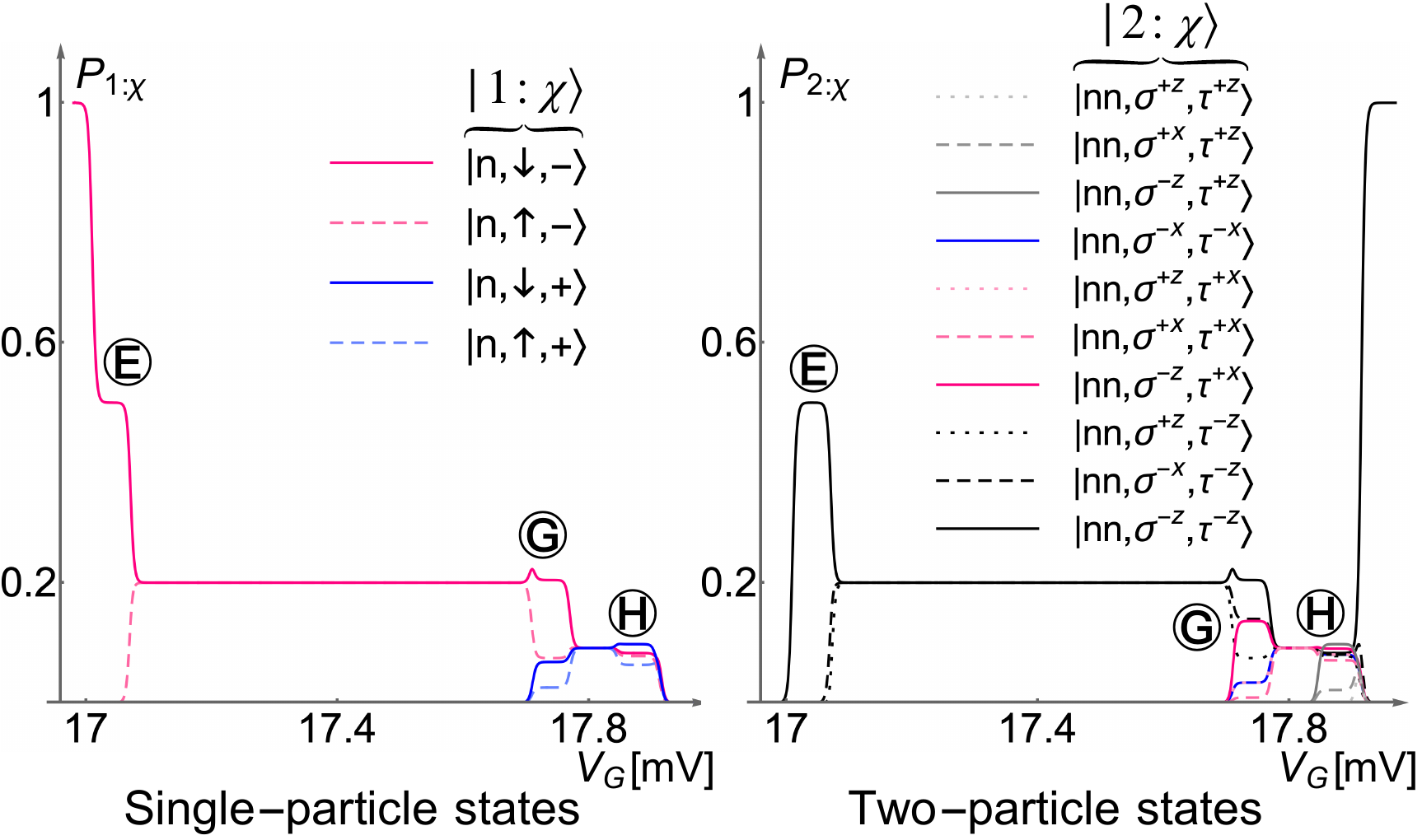}
\caption{Occupation probabilities for the \SP{} and  orbitally antisymmetric \TP{} states for cuts at $B=0.5$ T  along the gate voltage axis through the symmetrically coupled differential conductance maps in \fig\ref{fig:StrongMain} in the sequential tunnelling regime.}
\label{fig:StrongPop}
\end{figure}
are absent in the sequential tunnelling differential conductance maps  as the $K^{+}$ \SP{} states are not populated at the values of gate voltage needed for these transitions. Electrons cannot reach the $K^{+}$ \SP{} states because all transitions lower in gate voltage, including the \GS{}-to-\GS{} transition $|\SPorbital,\downarrow,-\rangle\rightarrow \StatenmmminusOneTminusZ$, occur between valley $K^{-} $ polarized states. Cotunnelling transitions, when relevant, enable the transitions in \eqn\eqref{eqn:AddTrans}, by populating the $K^{+}$ \SP{} states via inelastic cotunnelling $|\SPorbital,\downarrow,-\rangle\rightarrow |\SPorbital,\uparrow,+\rangle, \;|\SPorbital,\downarrow,+\rangle$. This cotunnelling-induced repopulation makes sequential tunnelling from the $K^{+}$ \SP{} states possible leading to weak features in the differential conductance maps at the gate voltages required for the transitions in \eqn\eqref{eqn:AddTrans} (bottom row of \fig\ref{fig:StrongMain}). Additionally, we observe cotunnelling-induced transport features outside the Coulomb diamond similar to the case of the orbitally symmetric multiplet, \sect\ref{sec:S}. Similarly, all cotunnelling features are suppressed by magnetic field and asymmetric coupling to the leads. 
Figure \ref{fig:StrongPop} demonstrates the cotunnelling-mediated redistribution of electrons among the states by comparing the occupation probabilities in the purely sequential tunnelling and sequential tunnelling + cotunnelling regimes. We note that a finite \SO{} coupling gap $\Delta_{SO}$ leads to two split copies of fanning lines in the differential conductance maps as those in \fig\ref{fig:StrongMain}.


\subsection{Interplay of ground- and excited \TP{} state multiplets}
\label{sec:Cross}

 \begin{figure*}[t!]
 \centering
\includegraphics[width=0.7\linewidth]{
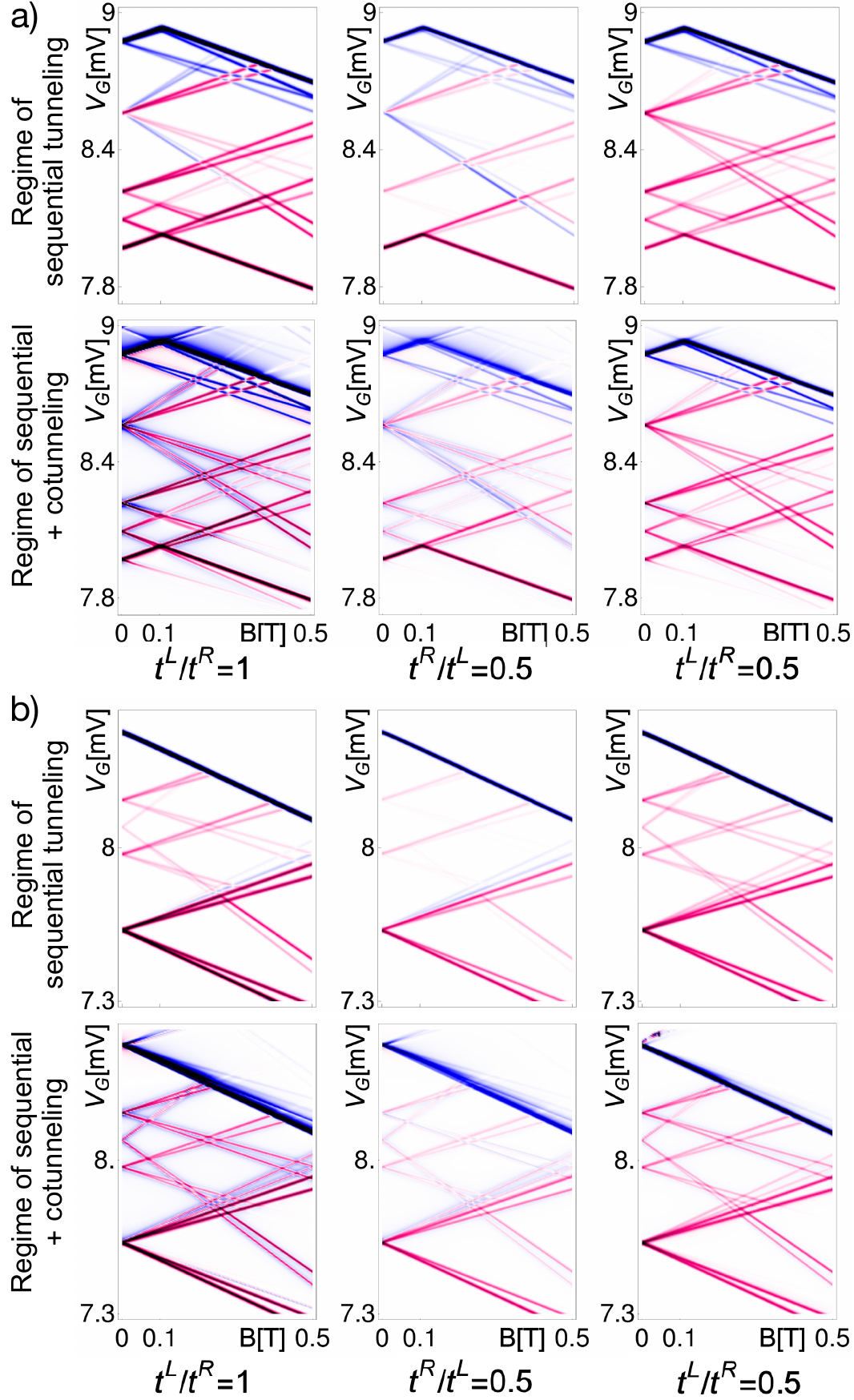}
\caption{Differential conductance maps for tunnelling transport through a \BLG{} \QD{} in which the \TP{} ground- and first excited state  are close enough in energy for both to be reached within the bias window are not merely superpositions of the two maps in \figs \ref{fig:WeakMain} and \ref{fig:StrongMain} for the two states due to interplay of the different multiplet states' occupation numbers. a) Orbitally symmetric \TP{} \GS{} and antisymmetric excited state, b) Orbitally antisymmetric \TP{} \GS{} and symmetric excited state. Parameters as in \fig\ref{fig:WeakMain}. }
\label{fig:MixedMain}
\end{figure*}

The dot's \TP{} ground and first excited state can be sufficiently close in energy for both to contribute transport signatures within the bias window\cite{knotheQuartetStatesTwoelectron2020}. We consider the cases in which the \TP{} \GS{} is either orbitally symmetric or antisymmetric, while the first excited state's orbital wave function is of the opposite symmetry. These scenarios yield distinct cases compared to the isolated \TP{} \GS{}s discussed in the previous sections. The \GS{}-to-\GS{} transitions originating from the \TP{} states of opposite symmetry can enable different transitions in the excited state multiplet compared to the isolated case. Also,  we can clearly distinguish the orbitally symmetric and antisymmetric \TP{} states by their zero-field splittings or absence thereof. Hence, investigating both simultaneously reveals changes in the orbital composition when comparing the ground and excited \TP{} states. 

Figure \ref{fig:MixedMain}a) depicts the differential conductance across a dot with an orbitally symmetric \TP{} \GS{} and orbitally antisymmetric first excited state. Transitions to both \TP{} multiplets manifest in the differential conductance maps. Notably, tunnelling channels involving valley-coherent \TP{} states in the orbitally symmetric \TP{} manifold lead to a population of the $K^{+}$ valley at sufficiently high gate voltages. These populations enable transitions to all the orbitally antisymmetric \TP{} states by purely sequential tunnelling. The differential conductance lines originating from either multiplet in \fig \ref{fig:MixedMain}a) have distinct slopes with $B$ due to the different orbital composition of the symmetric and antisymmetric orbital \TP{} wave functions yielding different valley g-factors. Since the orbital composition is unequal also for ground and excited states, the valley g-factors differ for the orbitally antisymmetric states in \fig\ref{fig:MixedMain}a) and \fig\ref{fig:StrongMain}. 


{Similar statements apply to the case of an orbitally antisymmetric \TP{} \GS{} and orbitally symmetric first excited state, \fig\ref{fig:MixedMain}b). Also here, an orbitally symmetric state occupies different orbitals, $n$, when being an excited state compared to a \GS{}, leading to different valley g-factors and different magnetic field splittings compared to \fig\ref{fig:WeakMain}. }

\section{Discussion and Conclusion}
\label{sec:conc}
In summary, we have analysed quantum tunnelling across an electrostatically induced \BLG{} \QD{} as a spectroscopic tool to resolve the dot's single and highly degenerate two-electron multiplets. Here, we summarise how to use tunnelling transport maps as a function of gate voltage and magnetic field to distinguish  the interaction regimes specified in \fig\ref{fig:Structure} and identify \TP{} states with different orbital, spin, and valley compositions:
\begin{itemize}
\item \textbf{The number and the splittings of peaks in the differential conductance at zero magnetic field tell about the orbital symmetry of the \TP{} wave function.} An orbitally antisymmetric \TP{} state (as for dots with weak screening and strong long-range Coulomb interaction, cf.~\fig\ref{fig:Structure}) hosts a tenfold degenerate multiplet of spin and valley states at $B=0$ (cf.~\fig\ref{fig:ASSplittings}), manifesting in one single transition. Conversely, the six possible spin and valley states of an  orbitally symmetric \TP{} state (which forms for strongly screened long-range interactions) are slightly split by short-range lattice-scale interactions (cf.~\fig\ref{fig:SSplittings}). Such splittings manifest in multiple possible transitions and corresponding tunnelling transport features within the bias window at zero magnetic field (\figs\ref{fig:SAll}, \ref{fig:WeakMain}, and \ref{fig:StrongMain}). 
\item \textbf{The various spin and valley states couple differently to a perpendicular magnetic field. Hence a magnetic field allows us to identify them and infer their g-factors.} Spin- and valley-polarized states split with $B$, while spin- and valley-coherent \TP{} states do not couple to a magnetic field. In combination with spin and valley selection rules, this contrasting magnetic field dependence helps identify the dispersing lines in the magnetic field-dependent differential conductance maps with the corresponding \SP{}-to-\TP{} transitions. The slope of these lines is proportional to the difference of the \SP{} and \TP{} states' g-factors. The orbital magnetic moment induced valley g-factor being much larger than the free particle spin g-factor allows distinguishing spin and valley splittings. 
\item \textbf{If multiple \TP{} states can be reached within the bias window, their distinct valley g-factors will help identify them.} The valley g-factor depends on the orbital wave function and its distribution in momentum space. Hence transitions from the same \SP{} state to distinct \TP{} states  show as lines with different slopes in the differential conductance maps, as in \fig\ref{fig:MixedMain}.
\end{itemize}

Our results will help to explain tunnelling transport experiments in \BLG{} \QD{}s in the one- and two-particle sectors. Identifying and controlling few-electron states is a crucial step towards using their degrees of freedom for quantum information storage and processing in future devices.

\section{Acknowledgements}
We acknowledge fruitful discussions with Luca Banszerus, Samuel Möller, Corinne Steiner, Eike Icking, Christian Volk, Christoph Stampfer,  Annika Kurzmann,  Chuyao Tong, Rebekka Garreis, Thomas Ihn, and Klaus Ensslin. VIF and AK were supported by EC-FET Core 3 European Graphene Flagship Project, EC-FET Quantum Flagship Project 2D-SIPC, and Lloyd Register Foundation Nanotechnology Grant. LIG was supported by the NSF DMR-2002275.  



\newpage
\appendix
\section{Sequential tunnelling rates}
\label{app:SeqTunRates}
\begin{widetext}
The sequential tunnelling rates, $W^{l}_{2:\chi\leftarrow1:\chi^{\prime}}$, in \eqn\eqref{eqn:WSeqT} for transitions from \SP{} dot levels to the orbitally symmetric \TP{} states, \eqn\eqref{eqn:sStates}, are given by,
\begin{align}
\nonumber & W^{l}_{\LnnSzeroTplusX\leftarrow \SPorbital\downminus}=\frac{2\pi}{\hbar}|t^{l}_{\SPorbital\upplus}|^{2}f(E_{\LnnSzeroTplusX}-E_{\SPorbital\downminus}-\Chempot^{l}),\\
\nonumber & W^{l}_{\LnnSzeroTplusX\leftarrow\SPorbital\upminus }=\frac{2\pi}{\hbar}|t^{l}_{\SPorbital\downplus}|^{2}f(E_{\LnnSzeroTplusX}-E_{\SPorbital\upminus}-\Chempot^{l}),\\
\nonumber & W^{l}_{\LnnSzeroTplusX\leftarrow\SPorbital\downplus }=\frac{2\pi}{\hbar}|t^{l}_{\SPorbital\upminus}|^{2}f(E_{\LnnSzeroTplusX}-E_{\SPorbital\downplus}-\Chempot^{l}),\\
\nonumber & W^{l}_{\LnnSzeroTplusX\leftarrow\SPorbital\upplus }=\frac{2\pi}{\hbar}|t^{l}_{\SPorbital\downminus}|^{2}f(E_{\LnnSzeroTplusX}-E_{\SPorbital\upplus}-\Chempot^{l}),\\
 \nonumber & W^{l}_{\LnnSzeroTminusZ\leftarrow\SPorbital\downminus }=\frac{2\pi}{\hbar}|t^{l}_{\SPorbital\upminus}|^{2}f(E_{\LnnSzeroTminusZ}-E_{\SPorbital\downminus}-\Chempot^{l}),\\
  \nonumber & W^{l}_{\LnnSzeroTminusZ\leftarrow\SPorbital\upminus }=\frac{2\pi}{\hbar}|t^{l}_{\SPorbital\downminus}|^{2}f(E_{\LnnSzeroTminusZ}-E_{\SPorbital\upminus}-\Chempot^{l}),\\
 \nonumber & W^{l}_{\LnnSzeroTminusZ\leftarrow\SPorbital\downplus }
= W^{l}_{\LnnSzeroTminusZ\leftarrow\SPorbital\upplus }
 = W^{l}_{\LnnSzeroTplusZ\leftarrow\SPorbital\downminus }
=W^{l}_{\LnnSzeroTplusZ\leftarrow\SPorbital\upminus }=0,\\
  \nonumber & W^{l}_{\LnnSzeroTplusZ\leftarrow\SPorbital\downplus }=\frac{2\pi}{\hbar}|t^{l}_{\SPorbital\upplus}|^{2}f(E_{\LnnSzeroTplusZ}-E_{\SPorbital\downplus}-\Chempot^{l}),\\
 \nonumber & W^{l}_{\LnnSzeroTplusZ\leftarrow\SPorbital\upplus }=\frac{2\pi}{\hbar}|t^{l}_{\SPorbital\downplus}|^{2}f(E_{\LnnSzeroTplusZ}-E_{\SPorbital\upplus}-\Chempot^{l}),\\
 \nonumber & W^{l}_{\LnnmminusOneTminusX\leftarrow\SPorbital\downminus }=\frac{2\pi}{\hbar}|t^{l}_{\SPorbital\downplus}|^{2}f(E_{\LnnmminusOneTminusX}-E_{\SPorbital\downminus}-\Chempot^{l}),\\
\nonumber & W^{l}_{\LnnmminusOneTminusX\leftarrow\SPorbital\upminus }=0,\\
 \nonumber & W^{l}_{\LnnmminusOneTminusX\leftarrow\SPorbital\downplus }=\frac{2\pi}{\hbar}|t^{l}_{\SPorbital\downminus}|^{2}f(E_{\LnnmminusOneTminusX}-E_{\SPorbital\downplus}-\Chempot^{l}),\\  
\nonumber & W^{l}_{\LnnmminusOneTminusX\leftarrow\SPorbital\upplus }
= W^{l}_{\LnnmplusOneTminusX\leftarrow\SPorbital\downminus }=0,\\      
 \nonumber & W^{l}_{\LnnmplusOneTminusX\leftarrow\SPorbital\upminus }=\frac{2\pi}{\hbar}|t^{l}_{\SPorbital\upplus}|^{2}f(E_{\LnnmplusOneTminusX}-E_{\SPorbital\upminus}-\Chempot^{l}),\\    
 \nonumber & W^{l}_{\LnnmplusOneTminusX\leftarrow\SPorbital\downplus }=0,\\       
 \nonumber & W^{l}_{\LnnmplusOneTminusX\leftarrow\SPorbital\upplus }=\frac{2\pi}{\hbar}|t^{l}_{\SPorbital\upminus}|^{2}f(E_{\LnnmplusOneTminusX}-E_{\SPorbital\upplus}-\Chempot^{l}),\\  
 \nonumber & W^{l}_{\LnnmZeroTminusX\leftarrow\SPorbital\downminus }=\frac{2\pi}{\hbar}|t^{l}_{\SPorbital\upplus}|^{2}f(E_{\LnnmZeroTminusX}-E_{\SPorbital\downminus}-\Chempot^{l}),\\
  \nonumber & W^{l}_{\LnnmZeroTminusX\leftarrow\SPorbital\upminus }=\frac{2\pi}{\hbar}|t^{l}_{\SPorbital\downplus}|^{2}f(E_{\LnnmZeroTminusX}-E_{\SPorbital\upminus}-\Chempot^{l}),\\
 \nonumber & W^{l}_{\LnnmZeroTminusX\leftarrow\SPorbital\downplus }=\frac{2\pi}{\hbar}|t^{l}_{\SPorbital\upminus}|^{2}f(E_{\LnnmZeroTminusX}-E_{\SPorbital\downplus}-\Chempot^{l}),\\
 & W^{l}_{\LnnmZeroTminusX\leftarrow\SPorbital\upplus }=\frac{2\pi}{\hbar}|t^{l}_{\SPorbital\downminus}|^{2}f(E_{\LnnmZeroTminusX}-E_{\SPorbital\upplus}-\Chempot^{l}),
\end{align}
in terms of the tunnelling amplitudes, $t^{l}_{\xi}$, and chemical potential, $\Chempot^{l}$, of the left ($l=L$) and right ($l=R$) lead. 
\end{widetext}

\begin{widetext}
The sequential tunnelling rates involving the orbitally antisymmetric \TP{} states, \eqn\eqref{eqn:sStates}, read,
\begin{align}
\nonumber & W^{l}_{\LnmmminusOneTminusZ\leftarrow\SPorbital\downminus }=\frac{2\pi}{\hbar}|t^{l}_{\SPorbital\downminus}|^{2}f(E_{\LnmmminusOneTminusZ}-E_{\SPorbital\downminus}-\Chempot^{l})\\
\nonumber & W^{l}_{\LnmmminusOneTminusZ\leftarrow\SPorbital\upminus }
= W^{l}_{\LnmmminusOneTminusZ\leftarrow\SPorbital\downplus }
= W^{l}_{\LnmmminusOneTminusZ\leftarrow\SPorbital\upplus }=0\\
\nonumber & W^{l}_{\LnmmZeroTminusZ\leftarrow\SPorbital\downminus }=\frac{2\pi}{\hbar}|t^{l}_{\SPorbital\upminus}|^{2}f(E_{\LnmmZeroTminusZ}-E_{\SPorbital\downminus}-\Chempot^{l})\\
\nonumber & W^{l}_{\LnmmZeroTminusZ\leftarrow\SPorbital\upminus }=\frac{2\pi}{\hbar}|t^{l}_{\SPorbital\downminus}|^{2}f(E_{\LnmmZeroTminusZ}-E_{\SPorbital\upminus}-\Chempot^{l})\\
\nonumber & W^{l}_{\LnmmZeroTminusZ\leftarrow\SPorbital\downplus }
=W^{l}_{\LnmmZeroTminusZ\leftarrow\SPorbital\upplus }
=W^{l}_{\LnmmplusOneTminusZ\leftarrow\SPorbital\downminus }=0\\
\nonumber & W^{l}_{\LnmmplusOneTminusZ\leftarrow\SPorbital\upminus }=\frac{2\pi}{\hbar}|t^{l}_{\SPorbital\upminus}|^{2}f(E_{\LnmmplusOneTminusZ}-E_{\SPorbital\upminus}-\Chempot^{l})\\
\nonumber & W^{l}_{\LnmmplusOneTminusZ\leftarrow\SPorbital\downplus }
= W^{l}_{\LnmmplusOneTminusZ\leftarrow|\SPorbital\upplus }=0\\
\nonumber & W^{l}_{\LnmmminusOneTplusX\leftarrow\SPorbital\downminus }=\frac{2\pi}{\hbar}|t^{l}_{\SPorbital\downplus}|^{2}f(E_{\LnmmminusOneTplusX}-E_{\SPorbital\downminus}-\Chempot^{l})\\
\nonumber & W^{l}_{\LnmmminusOneTplusX\leftarrow\SPorbital\upminus }=0\\
\nonumber & W^{l}_{\LnmmminusOneTplusX\leftarrow\SPorbital\downplus }=\frac{2\pi}{\hbar}|t^{l}_{\SPorbital\downminus}|^{2}f(E_{\LnmmminusOneTplusX}-E_{\SPorbital\downplus}-\Chempot^{l})\\
\nonumber & W^{l}_{\LnmmminusOneTplusX\leftarrow\SPorbital\upplus }=0\\
\nonumber & W^{l}_{\LnmmZeroTminusZ\leftarrow\SPorbital\downminus }=\frac{2\pi}{\hbar}|t^{l}_{\SPorbital\upplus}|^{2}f(E_{\LnmmZeroTminusZ}-E_{\SPorbital\downminus}-\Chempot^{l})\\
\nonumber & W^{l}_{\LnmmZeroTminusZ\leftarrow\SPorbital\upminus }=\frac{2\pi}{\hbar}|t^{l}_{\SPorbital\downplus}|^{2}f(E_{\LnmmZeroTminusZ}-E_{\SPorbital\upminus}-\Chempot^{l})\\
\nonumber & W^{l}_{\LnmmZeroTminusZ\leftarrow\SPorbital\downplus }=\frac{2\pi}{\hbar}|t^{l}_{\SPorbital\upminus}|^{2}f(E_{\LnmmZeroTminusZ}-E_{\SPorbital\downplus}-\Chempot^{l})\\
\nonumber & W^{l}_{\LnmmZeroTminusZ\leftarrow\SPorbital\upplus }=\frac{2\pi}{\hbar}|t^{l}_{\SPorbital\downminus}|^{2}f(E_{\LnmmZeroTminusZ}-E_{\SPorbital\upplus}-\Chempot^{l})\\
\nonumber & W^{l}_{\LnmmplusOneTminusZ\leftarrow\SPorbital\downminus }=0\\
\nonumber & W^{l}_{\LnmmplusOneTminusZ\leftarrow\SPorbital\upminus }=\frac{2\pi}{\hbar}|t^{l}_{\SPorbital\upplus}|^{2}f(E_{\LnmmplusOneTminusZ}-E_{\SPorbital\upminus}-\Chempot^{l})\\
\nonumber & W^{l}_{\LnmmplusOneTminusZ\leftarrow\SPorbital\downplus }=0\\
\nonumber & W^{l}_{\LnmmplusOneTminusZ\leftarrow\SPorbital\upplus }=\frac{2\pi}{\hbar}|t^{l}_{\SPorbital\upminus}|^{2}f(E_{\LnmmplusOneTminusZ}-E_{\SPorbital\upplus}-\Chempot^{l})\\
\nonumber & W^{l}_{\LnmSzeroTminusX\leftarrow\SPorbital\downminus }=\frac{2\pi}{\hbar}|t^{l}_{\SPorbital\upplus}|^{2}f(E_{\LnmSzeroTminusX}-E_{\SPorbital\downminus}-\Chempot^{l})\\
\nonumber & W^{l}_{\LnmSzeroTminusX\leftarrow\SPorbital\upminus }=\frac{2\pi}{\hbar}|t^{l}_{\SPorbital\downplus}|^{2}f(E_{\LnmSzeroTminusX}-E_{\SPorbital\upminus}-\Chempot^{l})\\
\nonumber & W^{l}_{\LnmSzeroTminusX\leftarrow\SPorbital\downplus }=\frac{2\pi}{\hbar}|t^{l}_{\SPorbital\upminus}|^{2}f(E_{\LnmSzeroTminusX}-E_{\SPorbital\downplus}-\Chempot^{l})\\
\nonumber & W^{l}_{\LnmSzeroTminusX\leftarrow\SPorbital\upplus }=\frac{2\pi}{\hbar}|t^{l}_{\SPorbital\downminus}|^{2}f(E_{\LnmSzeroTminusX}-E_{\SPorbital\upplus}-\Chempot^{l})\\
\nonumber & W^{l}_{\LnmmminusOneTplusZ\leftarrow\SPorbital\downminus }
= W^{l}_{\LnmmminusOneTplusZ\leftarrow\SPorbital\upminus }=0\\
\nonumber & W^{l}_{\LnmmminusOneTplusZ\leftarrow\SPorbital\downplus }=\frac{2\pi}{\hbar}|t^{l}_{\SPorbital\downplus}|^{2}f(E_{\LnmmminusOneTplusZ}-E_{\SPorbital\downplus}-\Chempot^{l})\\
\nonumber & W^{l}_{\LnmmminusOneTplusZ\leftarrow\SPorbital\upplus }
= W^{l}_{\LnmmZeroTplusZ\leftarrow\SPorbital\downminus }
= W^{l}_{\LnmmZeroTplusZ\leftarrow\SPorbital\upminus }=0\\
\nonumber & W^{l}_{\LnmmZeroTplusZ\leftarrow\SPorbital\downplus }=\frac{2\pi}{\hbar}|t^{l}_{\SPorbital\upplus}|^{2}f(E_{\LnmmZeroTplusZ}-E_{\SPorbital\downplus}-\Chempot^{l})\\
\nonumber & W^{l}_{\LnmmZeroTplusZ\leftarrow\SPorbital\upplus }=\frac{2\pi}{\hbar}|t^{l}_{\SPorbital\downplus}|^{2}f(E_{\LnmmZeroTplusZ}-E_{\SPorbital\upplus}-\Chempot^{l})\\
\nonumber & W^{l}_{\LnmmplusOneTplusZ\leftarrow\SPorbital\downminus }
= W^{l}_{\LnmmplusOneTplusZ\leftarrow\SPorbital\upminus }
= W^{l}_{\LnmmplusOneTplusZ\leftarrow\SPorbital\downplus }=0\\
 & W^{l}_{\LnmmplusOneTplusZ\leftarrow\SPorbital\upplus }=\frac{2\pi}{\hbar}|t^{l}_{\SPorbital\upplus}|^{2}f(E_{\LnmmplusOneTplusZ}-E_{\SPorbital\upplus}-\Chempot^{l}).
\end{align}
\end{widetext}

\section{Cotunnelling rates}
\label{app:CoTunRates}

By the regularization scheme described in the main text, the cotunnelling rates, \eqn\eqref{eqn:WCoT}, evaluate to,
\begin{widetext}
\begin{align}
 \nonumber& W_{ 1:\chi\leftarrow1:\chi^{\prime}}=\sum_{l,\l^{\prime}}  \nonumber W^{l,\l^{\prime}}_{ 1:\chi\leftarrow1:\chi^{\prime}}\\
\nonumber &=\frac{2\pi}{\hbar}\sum_{l,\l^{\prime},\tilde\chi}|t_{\chi^{\prime}}^{l}\; t_{\chi}^{l^{\prime}}\hskip0pt^*|^2 \iint d\epsilon_{k}^{l^{\prime}}d\epsilon_{k^{\prime}}^{l}\Big| \frac{1}{E_{i,1:\chi^{\prime}}-E_{2:{\tilde{\chi}}}+\epsilon_{k}^{l^{\prime}}+i\gamma}    \Big|^{2} f(\epsilon_{k}^{l^{\prime}}-\Chempot^{l^{\prime}} ) \big[ 1- f(\epsilon_{k^{\prime}}^{l}-\Chempot^{l} ) \big] \delta (E_{1:\chi} + \epsilon_{k^{\prime}}^{l}-E_{1:\chi^{\prime}}-\epsilon_{k}^{l^{\prime}}) \\
 &=\frac{2\pi}{\hbar}\sum_{l,\l^{\prime},\tilde\chi}|t_{\chi^{\prime}}^{l}\; t_{\chi}^{l^{\prime}}\hskip0pt^*|^2   J\big(\Chempot^{l^{\prime}}, \Chempot^{l} +E_{1:\chi}-E_{1:\chi^{\prime}} , -E_{1:\chi^{\prime}}+E_{2 :{\tilde{\chi}}}\big), \\
 \nonumber &W_{2:\chi\leftarrow2:\chi^{\prime}}=\sum_{l,\l^{\prime}}   W^{l,\l^{\prime}}_{2:\chi\leftarrow2:\chi^{\prime}}\\
\nonumber &=\frac{2\pi}{\hbar}\sum_{l,\l^{\prime},\tilde\chi}|t_{\chi^{\prime}}^{l}\; t_{\chi}^{l^{\prime}}\hskip0pt^*|^2 \iint d\epsilon_{k}^{l^{\prime}}d\epsilon_{k^{\prime}}^{l}\Big|  \frac{1}{E_{i,2:\chi^{\prime}}-E_{1:\tilde{{\chi}}}-\epsilon_{k^{\prime}}^{l}+i\gamma}     \Big|^{2} f(\epsilon_{k}^{l^{\prime}}-\Chempot^{l^{\prime}} ) \big[ 1- f(\epsilon_{k^{\prime}}^{l}-\Chempot^{l} ) \big] \delta (E_{2:\chi} + \epsilon_{k^{\prime}}^{l}-E_{2:\chi^{\prime}}-\epsilon_{k}^{l^{\prime}}) \\
 &=\frac{2\pi}{\hbar}\sum_{l,\l^{\prime},\tilde\chi}|t_{\chi^{\prime}}^{l}\; t_{\chi}^{l^{\prime}}\hskip0pt^*|^2 
 J\big(\Chempot^{l^{\prime}} , \Chempot^{l} +E_{2:\chi}-E_{2:\chi^{\prime}} ,  -E_{1:\tilde{{\chi}}}+E_{2:\chi}\big),
 \end{align}
exploiting the relation
\begin{align}
\nonumber J(\mu_{1}, \mu_{2}, E) =& \lim_{\gamma\to0}\int d\epsilon f(\epsilon-\mu_{1}) \big[1- f(\epsilon-\mu_{2}) \big] \frac{1}{(\epsilon-E)^{2}+\gamma^{2}}-\mathcal{O}(\frac{1}{\gamma})\\
=&\frac{1}{2\pi k_{B}T} n_{B}(\mu_{2}-\mu_{1})\mathfrak{Im}\Big[\psi^{\prime}(\frac{1}{2}+i\frac{\mu_{2}-E}{2\pi k_{B}T})-\psi^{\prime}(\frac{1}{2}+i\frac{\mu_{1}-E}{2\pi k_{B}T})\Big],
\end{align}
in terms of the Bose function $n_{B}$ and the polygamma function $\psi$.
\end{widetext}

\section{Occupation probabilities}
\label{app:CoTunProps}

Similar to \figs \ref{fig:WeakPop} and \ref{fig:StrongPop} in the main text, here we discuss how the occupation probabilities of the \SP{} and the orbitally symmetric or the orbitally antisymmetric \TP{} states change when varying the gate voltage at fixed values of the bias voltage and the magnetic field. Figures \ref{fig:WeakPopAll} and \ref{fig:StrongPopAll} compare the occupation probabilities of the dot states for sequential tunnelling to the regime of sequential + cotunnelling for the two different \TP{} multiplets. We observe how the additional inter-multiplet transitions induced by cotunnelling processes alter the states' occupations and enable different tunnelling sequences compared to purely sequential tunnelling.
 \begin{figure*}[t!]
 \centering
\includegraphics[width=1\linewidth]{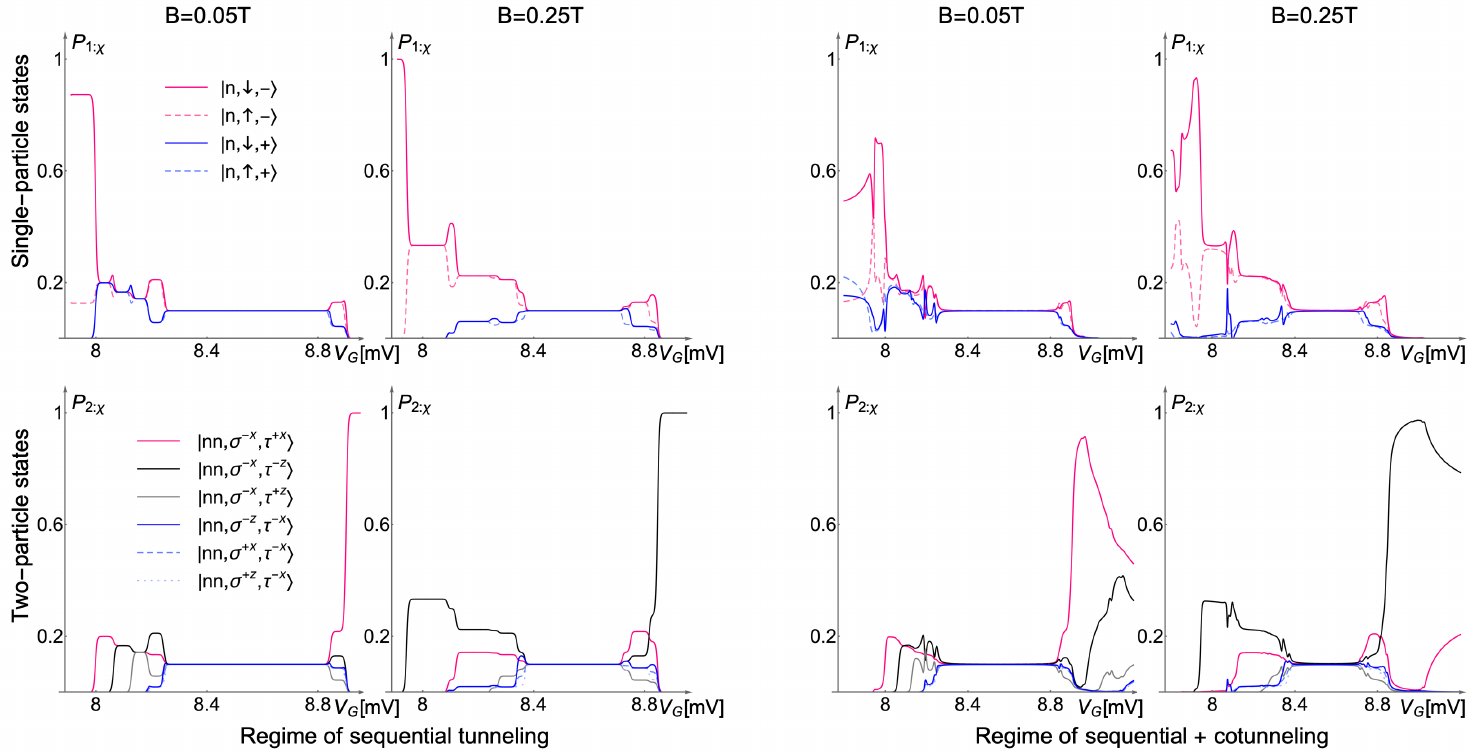}
\caption{{ Probabilities for the \SP{} and  orbitally symmetric \TP{} states to be occupied for fixed magnetic field cuts  along the gate voltage axis through the symmetrically coupled differential conductance maps with zero \SO{} coupling in \fig\ref{fig:WeakMain} (leftmost differential conductance maps). }}
\label{fig:WeakPopAll}
\end{figure*}

 \begin{figure}[t!]
 \centering
\includegraphics[width=1\linewidth]{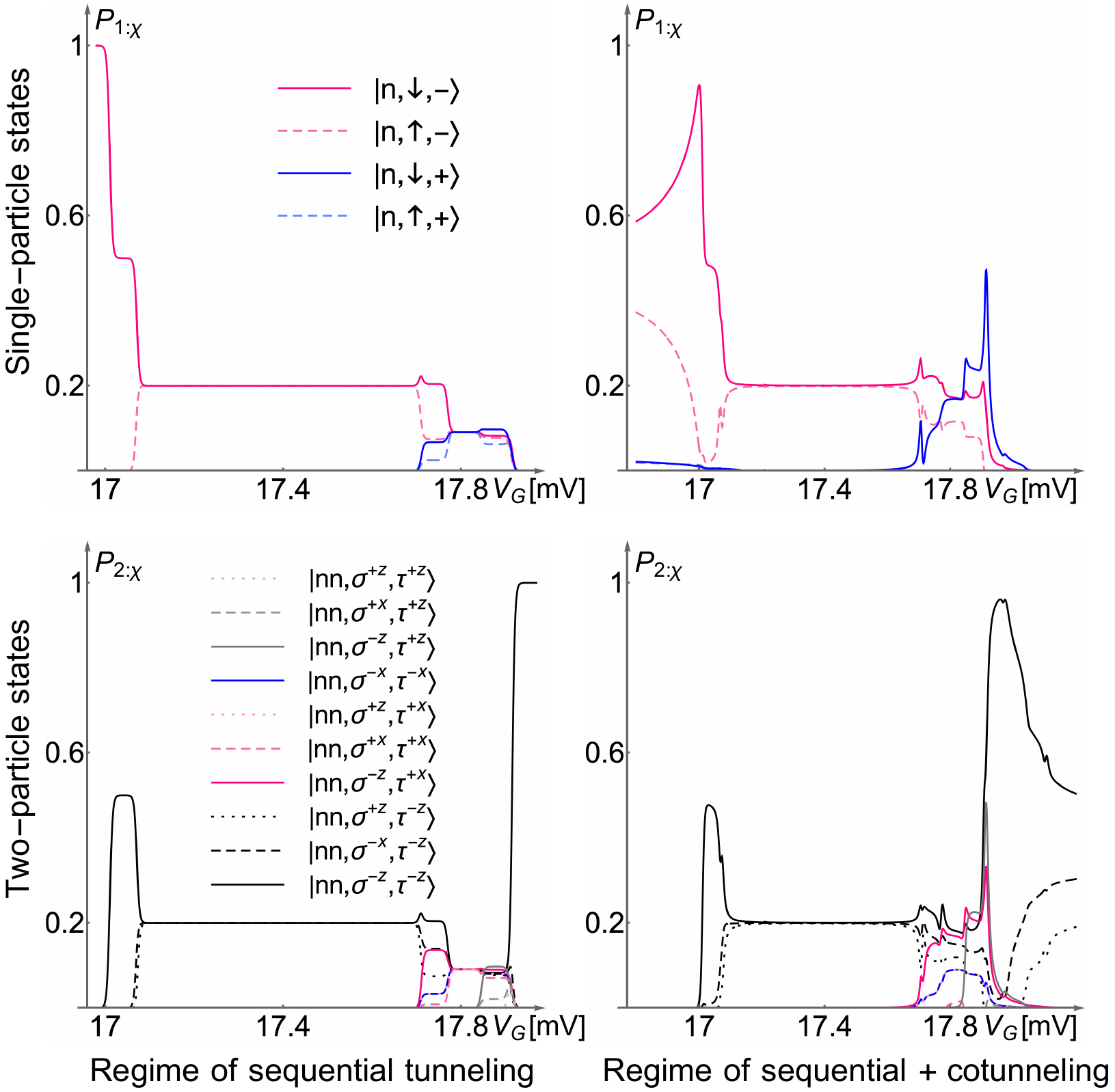}
\caption{Probabilities for the \SP{} and  orbitally antisymmetric \TP{} states to be occupied cuts at $B=0.5$ T  along the gate voltage axis through the symmetrically coupled differential conductance maps in \fig\ref{fig:StrongMain}. }
\label{fig:StrongPopAll}
\end{figure}

\bibliography{QD2}

\end{document}